\documentclass[journal,onecolumn]{IEEEtran}

\usepackage[T1]{fontenc}

\usepackage{cite}

\usepackage[cmintegrals]{newtxmath}

\renewcommand{\caption}[2][\relax]{\MYoriglatexcaption[#2]{#2}}

\usepackage{epsf}
\usepackage{graphicx}
\usepackage{amsmath}
\usepackage{epsfig,latexsym,epsf,amsfonts}
\usepackage{algorithm, algorithmic}
\usepackage{placeins}
\usepackage{array,color}
\usepackage{cite}
\usepackage{textcomp}
\usepackage{lipsum}

\newcommand{\cmmnt}[1]{\ignorespaces}

\def\baselinestretch{6.5} 
\normalsize 
\newcommand{\PublishMode}[1]{} 
\newcommand{\DraftMode}[1]{#1}  

\newcommand{\comment}[1]{}  
\DraftMode{
	\oddsidemargin  -0in
	\evensidemargin -0.2in
	\topmargin -0.4in
	\headheight 15truept
	\textheight 9.1in \textwidth 5.7in
	\linespread{2}  
}

\begin{document}
	
	\title{\huge{Neural Network Cognitive Engine for Autonomous and Distributed Underlay Dynamic Spectrum Access}}
	
	\author{Fatemeh Shah-Mohammadi and Andres Kwasinski\\
		Kate Gleason College of Engineering\\
		Rochester Institute of Technology, Rochester, New York 14623, USA\thanks{ }}
	\maketitle \thispagestyle{empty}
	
\begin{abstract}
Two key challenges in underlay dynamic spectrum access (DSA) are how to establish an interference limit from the primary network (PN) and how cognitive radios (CRs) in the secondary network (SN) become aware of the interference they create on the PN, especially when there is no exchange of information between the two networks. These challenges are addressed in this paper by presenting a fully autonomous and distributed underlay DSA scheme where each CR operates based on predicting its transmission effect on the PN. The scheme is based on a cognitive engine with an artificial neural network that predicts, without exchanging information between the networks, the adaptive modulation and coding configuration for the primary link nearest to a transmitting CR. By managing the effect of the SN on the PN, the presented technique maintains the relative average throughput change in the PN within a prescribed maximum value, while also finding transmit settings for the CRs that result in throughput as large as allowed by the PN interference limit. Simulation results show that the ability of the cognitive engine in estimating the effect of a CR transmission on the full adaptive modulation and coding (AMC) mode leads to a much more fine underlay transmit power control. This ability also provides higher transmission opportunities for the CRs, compared to a scheme that can only estimate the modulation scheme used at the PN link.

\end{abstract}

\section{Introduction}\label{sec:intro}

The cognitive radio (CR) paradigm is seen as a key solution to the radio spectrum scarcity problem stemming from the inefficiency of the established static spectrum allocation policy, \cite{force2002spectrum}, and the growing connectivity needs from wireless applications. A CR is a wireless device with the ability to autonomously gain awareness of the wireless network environment and to learn how to adapt its operating parameters to best meet the end-user goals \cite{haykin2005cognitive}. As a result, CRs have frequently been considered as an enabling technology for dynamic spectrum access (DSA). Through DSA, a network of CRs, called the secondary network (SN), operates by sharing the radio spectrum with a primary network (PN) which is the incumbent owner of the spectrum band in use. Of the three main DSA approaches (overlay, underlay, and interweave DSA \cite{b1}), this paper focuses on underlay DSA. The underlay DSA allows CRs in the SN (called hereafter ``secondary users'' - SUs) to transmit over the same spectrum band being used by the PN by limiting their transmit power level so that the interference they create on the PN remains below a tolerable threshold \cite{b2}.

Despite having been studied for almost two decades, the main challenges in underlay DSA remain on how to establish the interference threshold for the PN links and how the SUs autonomously become aware of the interference they create on the PN, specially when following an ideal operating setup where there is no exchange of information between primary and secondary networks. In order for the SN to assess its effect on the PN and protect the PN transmissions, researchers have proposed different techniques that usually assume that the secondary transmitter ($SU_{TX}$) knows the gain of the primary channel (that between the primary transmitter, $PU_{TX}$, and the primary receiver, $PU_{RX}$) and/or the cross-channel gain from the $SU_{TX}$ to the $PU_{RX}$, \cite{hu2009dynamic,b111,b112,b115,b116,b117,b118,b119,b120}. The common theme between these works is that they make use of the information that is sent over a feedback channel in the PN. Since feedback channels are part of most wireless communication standards \cite{3GPP1,3GPP2}, they have often been used, under the assumption that SUs can access them, to not only estimate the primary channel gain, but also assess the effect of the SN on PN transmissions. Examples of a CR obtaining information about the primary link from a PN feedback channel are found in \cite{b111,b112,b115,b116,b120} for the case of the rate/power control feedback channel, in \cite{b117} and \cite{b118} for the ARQ feedback channel, and in \cite{b119} for the feedback of the channel state information (CSI). In general, by relying on listening to feedback channels from the PN all works mentioned above share a setup where primary and secondary networks are not completely separated and exchange information with each other. In fact, the access of a control channel from another network calls into question as to whether there are really two separate network or, as we would argue, a single network with two different types of nodes. 

With a different approach, the works in \cite{b113} and \cite{b114} proposed solutions to obtain the cross-channel gain without listening to the PN feedback channels. The typical process in these works consists of the SU observing a change in the primary's waveform power and/or modulation order that results from the transmission of a probe message from the SU, \cite{b113}, or the SU acting as a relay by sending the amplified version of the signal received from the PU, \cite{b114}. However, in these works a CR transmitter can not estimate the cross-channel gain, and later assess the effect of SN on PN, unless it observes a change in primary signal power and/or modulation order. Moreover, these work did not considered the combined effect from scenarios with multiple links in the PN and the SN, as they focused on a setup with one link in each network. 
In contrast to these works, the technique to be presented here is not limited by the need to observe a change in transmit power or modulation order and is able to directly estimate the effect of an SN transmission on the PN much more accurately by also estimating the channel coding rate in a PN link. Moreover, the work herein is on scenarios consisting of multiple links in the PN and SN. 
In addition, our presented technique meets a key requirement by not relying on any information exchange between the networks. But, instead, our technique takes advantage of the use at the PN of adaptive modulation and coding (AMC), a technique where the modulation scheme and channel coding rate (a pair of parameters known as the AMC mode) are adapted based on the quality of the transmission link. The use of AMC has been part of all high performance wireless communications standards developed over the past two decades and, thus, expected to be used by a typical PN \cite{tarhini2007capacity,yang2002adaptive,zhou2018energy}. Since the AMC mode in use depends on the link's SINR, by estimating the AMC mode used in a primary link, it becomes possible for a CR to learn the signal-to-interference-plus-noise ratio (SINR) experienced at that link, from which the effect of CR transmission on the PU link can be assessed. Specifically, we propose an underlay DSA technique that configures the transmit power for a SU based on estimating the throughput at its nearest PN link that corresponds to the AMC mode that would be chosen based on the interference created by the SU's transmission.

On the other hand, leveraging the use of adaptive modulation in the PN will allow us not only to assess the effects of the SN on the PN, but will also allow us to establish the PN interference threshold. As shown in \cite{b5}, the use of adaptive modulation allows for the background noise to increase up to a certain level before the average throughput in a network starts to decrease. In the context of underlay DSA where the PN does not exchange any information with the SN, the interference imposed on the PN by an underlay-transmitting CR can be seen as a background noise for the PN, that can be increased up to a level which does not affect the average throughput in the primary network. Therefore, a CR can become aware about the interference that is creating on the primary link and decide on its transmit power by inferring the change in experienced throughput of the primary link (details in section \ref{AMC}). 
	

The core of our technique is a non-linear autoregressive exogenous neural network (NARX-NN)-based cognitive engine (CE) that estimates the throughput on a PN link (equivalently, the full AMC mode), an indicator of the interference induced by the SU on the PN link. The proposed CE uses as input an estimate of the modulation scheme being used at the PN link. As such, to the best of our knowledge, our presented cognitive engine is the first to be capable of estimating the channel coding rate setting for an AMC mode, a capability well beyond the estimation of the modulation scheme that has been realized through multiple signal processing or machine learning techniques. While modulation classification, the technique to estimate the modulation scheme used in a radio waveform, is applied in our DSA technique to derive an input for the cognitive engine, it is not the subject of this work. Instead, we leverage the already large volume of existing research. 
In this regard, the work in \cite{b27} studied modulation classification based on second and higher order time variant periodic cumulant function of the sensed signal for which it is required a prior knowledge of the signal parameters. Authors in \cite{b16} used the same framework to perform signal pre-processing, along with utilizing artificial neural networks to address the issues associated with classification when the signal parameters are unknown. The work in \cite{b28} proposed a fully automated modulation classification scheme which employs two stages of signal processing to classify the modulation of an incoming signal. In this paper we assume that each SU applies the technique proposed in \cite{b28} on the nearest primary link to infer its used modulation scheme (details are omitted here).

Therefore, the main contribution of this paper resides in presenting an underlay DSA technique to infer, without tapping into feedback or control channels from another network, the experienced throughput and accordingly the modulation order and the channel coding rate used in the transmission of the other network (the primary network here), and to leverage this inference in the realization of a fully autonomous and distributed underlay DSA scheme.
This ability to estimate the effective throughput enables a finer control knob for a more accurate power allocation at the SN with less harmful effect on PN transmissions as compared to techniques that rely only on the estimation of modulation order (from applying signal processing on the transmission waveform).

Our simulation results will show that the presented technique is able to maintain the relative change in PN average throughput within a prescribed fine-grained target maximum value (as an indicator of maximum allowed interference in PN), while at the same time finding transmit settings for the SUs that will result in as large throughput in the SN as could be allowed by the PN interference limit. As such, while succeeding in its main goal of autonomously and distributively determining the transmit power of the SUs such that the interference they create remains below the PN allowed interference limit, our proposed technique is also able to manage the tradeoff between the effect of the SN on the PN and the achievable throughput at the SN. Specifically, simulation results will show that for the proposed system with a target PN maximum relative average throughput change of 2\%, the achieved relative change is less than 3\%, while at the same time achieving useful average throughput values in the SN between 180 and 50 kbps. In addition, it will be seen that the implementation of a variation of the proposed scheme that reduces three times the overhead from transmitting probe messages still exhibits the ability to finely control the effect on the primary network throughput, although, as is to expect, compared to the case of sending all probe messages it increases by at most 1\%, only at the very low PN load of 0.16, the relative average throughput change. 
	
The rest of this paper is organized as follows. Section \ref{PNsetup} presents the overall system setup. Section \ref{MathAnalsis} presents an analytical framework to assess the effect of transmission in the SN on the PN, under the assumed fully autonomous transmission scenario. The rationale on how AMC can be leveraged to address the main two challenges associated with underlay DSA is outlined in Section \ref{AMC}. Section \ref{thetech} describes our proposed distributed underlay DSA technique. Simulation results are presented in Section \ref{results}, followed by conclusions in Section \ref{conclusion}.


\begin{figure}[tbp!]
		\centering
		\includegraphics[width=0.45\textwidth]{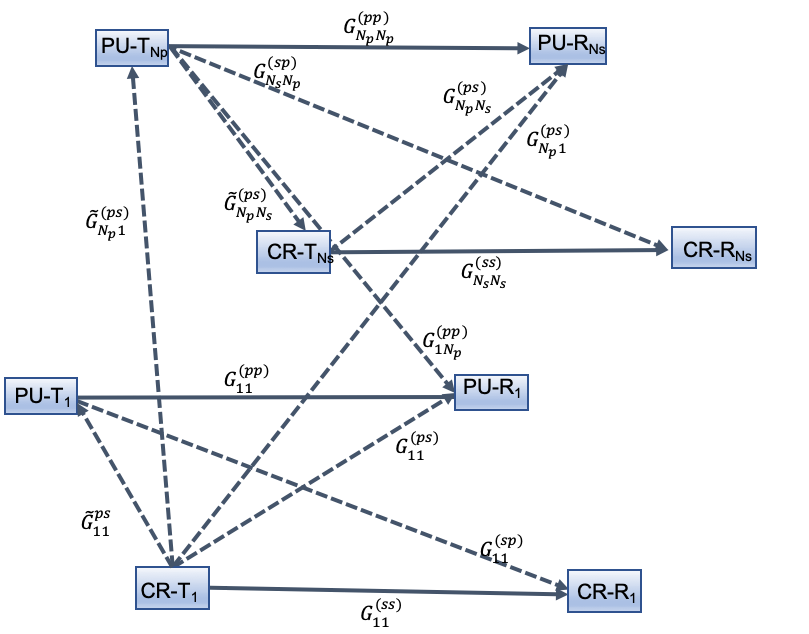}
		\caption{Considered network model composed of $N_p$ primary transceivers and $N_s$ cognitive radio users.}
		\label{sysmodel}
\end{figure}

\section{System Setup}
\label{PNsetup}
We consider a primary network with $N_P$ active primary links coexisting with $N_S$ active secondary links, with both networks transmitting over the same frequency band. The system model is shown in Fig.\ref{sysmodel}, where $G_{ij}^{(ps)}$ denoted as cross-channel gain is the path gain from $j$th. transmitting SU to the receiver in $i$th. primary link, $G_{ij}^{(ss)}$ is the path gain from the transmitter in the $j$th. secondary link to the receiver in the $i$th. secondary link, $G_{ij}^{(sp)}$ is the path gain from the transmitter in $j$th. primary link to the receiver in $i$th. SU link, $G_{ii}^{(pp)}$ is the path gain from the $i$th. primary transmitter to its corresponding receiver, and $G_{ii}^{(ss)}$ is the path gain from the $i$th. secondary transmitter to its corresponding receiver. While, $\Tilde{G}_{ij}^{(ps)}$ is the path gain from the transmitter in $j$th. secondary link to the transmitter in $i$th. primary link. 

In this section we focus on describing the operation of the PN, which is incumbent to the considered radio spectrum band. Section \ref{thetech} will present the underlay DSA scheme implemented in the SN. We assume that the PUs receive service from $N_{PBS}$ transmitting base stations (BSs), and we call the ratio $N_P/N_{PBS}$ as the primary network load. Each PU is assigned to the base station that presents the best channel gain. In addition, AMC is used in all transmissions (primary and secondary networks). This means that a transmitter has information about its link quality, in terms of SINR, and based on this assessment chooses, from a set of options, the modulation scheme and channel coding rate that results in highest throughput while at the same time meeting a maximum bit error rate (BER) limit.

Let $P_i^{(p)}$ denotes the transmit power in the $i$th. active primary link ($i=1,2,\dots,N_P$). Then, in the absence of the SN, the SINR in the $i$th. primary link (which is used to decide on the AMC mode) can be written as,
	\begin{eqnarray}
	\label{sinrprim}
	\gamma_i^{(p)} = \frac{G_{ii}^{(pp)} P_i^{(p)}}{\displaystyle{\sum_{j{\ne}i}} G_{ij}^{(pp)} P_j^{(p)} + \sigma_p^2 },{\quad}i=1,2,...,N_P,\\ [-.1in] \nonumber
\end{eqnarray}
where $\sigma_p^2$ is the background noise power.
	
In addition to AMC, without loss of generality, we adopt for the primary network the variable transmit power allocation algorithm proposed in \cite{b5}. This is an iterative power control algorithm that converges to a global optimum solution that maximizes the product of SINRs across all active links. In the algorithm, the transmit power at the $i$th. primary link is updated as,\vspace*{-2mm}
\begin{eqnarray}
\label{PrPowIter}
P_i^{(p)} \longleftarrow \left(\displaystyle{\sum_{j{\ne}i}} \: \frac{G_{ji}^{(pp)}}{\sum_{m{\ne}j} G_{jm}^{(pp)} P_m^{(p)} + \sigma_p^2}\right)^{-1}.\\ [-.1in] \nonumber
\end{eqnarray}

\section{SU Transmission Effect Assessment on the PU: Analytical Framework}\label{MathAnalsis}

In this section, we develop an analytical framework to explore, under a fully autonomous and distributed transmission scenario, the possibility of closed-form analytical estimation of the interference created by a transmitting CR to its nearest PU link. It is assumed that the PUs are
oblivious to the existence of the SUs and treat the interference
from transmitting CRs as additional noise at their receiver. As mentioned earlier, we consider a practical scenario where the CRs have no access to any control channel in the PN from which to obtain any side information (e.g., $G_{ij}^{(ps)}$) that might facilitate their interference control on the PN.

Similar to our proposed technique to be presented inhere, we assume a two-stage operation in the SN. During the first stage, every SU in the SN listens to the transmissions in the PN and observes the received signal power. In the second stage, every SU broadcasts a probing signal. This probing signal, transmitted with the same power by all SUs, will create some interfere on the PN. Since PUs make use of variable transmit power and AMC, upon experiencing the interference from the probing signal, every receiver in the PN sends back a control signal to its corresponding transmitter to adapt its transmit power and rate (specifically, the AMC mode) accordingly. These adaptation in transmission parameter at the primary links are observed by the SUs. We assume that SUs know the primary transmission protocol and are able to synchronize their operation with the primary transmissions, and that all the channel gains involved in Fig. \ref{sysmodel} remain constant during this process.

Let $P_{i}^{(p)}[0]$ denotes the initial transmit power in the $i$th. active primary link ($i=1,2,\dots,N_P$) while SUs listen to the transmissions in the PN. The received signal power at this stage (i.e. listening stage) at $i$th. SU transmitter can be written as $S_i[0]= \displaystyle{\sum_{j=1}^{N_P}}\Tilde{G}_{ij}^{(ps)}P_{j}^{(p)}[0]$. This equation further can be separated into two parts: one for the power received from the nearest PU transmitter and a second component for the received power from the rest of active PUs in the system, as in, 
\begin{eqnarray}
 S_i[0]= \Tilde{G}_{in}^{(ps)}P_{n}^{(p)}[0] +\displaystyle{\sum_{\substack{j=1 \\ j{\ne}n}}^{N_P}}\Tilde{G}_{ij}^{(ps)}P_{j}^{(p)}[0],\\ [-.2in] \nonumber  
\end{eqnarray}
 \noindent where $\Tilde{G}_{in}^{(ps)}P_{n}^{(p)}[0]$ shows the power received from the nearest PU active link (the subscript $n$ is used to highlight, among all active transmission links in the PN, the nearest link to the $i$th. SU).
Considering the same probing signal power for all SUs to be as $P^{(s)}$, the observed power at the $i$th. SU transmitter during the probing stage can be also written as:
\begin{eqnarray}
  S_i[1]= \Tilde{G}_{in}^{(ps)}P_{n}^{(p)}[1] +\displaystyle{\sum_{\substack{j=1 \\ j{\ne}n}}^{N_P}}\Tilde{G}_{ij}^{(ps)}P_{j}^{(p)}[1]+I_{ss},\\ [-.1in] \nonumber     
 \end{eqnarray}
\noindent where $I_{ss} = \displaystyle{\sum_{j=1,j{\ne}i}^{N_S}} G_{ij}^{(ss)} P^{(s)}$ denotes the total interference from the SN to the $i$th. transmitting SU, and $P_{j}^{(p)}[1]$ denotes, for the $j$th. PU transmit link, the new adapted transmit power. Without loss of generality it can be assumed that $P_{i}^{(p)}[0]>0$, and thus $S_i[0]>0 $, because if $P_{i}^{(p)}[0]=0$, the SU can simply transmits as the spectrum band is unoccupied; and the estimation of the created interference becomes unnecessary. The received signal powers during the listening and probing stage contain the information about the cross-channel gain between $i$th. SU transmitter and its corresponding nearest PU receiver $G_{in}^{(ps)}$. As mentioned earlier, the cross-channel gain can be used at the SU to assess its effect of transmission on the PU. Thus, in the following we explore whether each SU can determine $G_{in}^{(ps)}$ using its observations. By dividing the signal powers received at the $i$th. SU transmitter across two listening and probing stages $S_i[0]$ and $S_i[1]$, we have:
 
\begin{eqnarray}\label{probing}
\frac{S_i[1]}{S_i[0]}= \frac{\Tilde{G}_{in}^{(ps)}P_{n}^{(p)}[1] +\displaystyle{\sum_{j{\ne}n}^{N_P}}\Tilde{G}_{ij}^{(ps)}P_{j}^{(p)}[1]+I_{ss}}{\Tilde{G}_{in}^{(ps)}P_{n}^{(p)}[0] +\displaystyle{\sum_{j=1,j{\ne}n}^{N_P}}\Tilde{G}_{ij}^{(ps)}P_{j}^{(p)}[0]}
\end{eqnarray}

On the other hand, based on the assumption we made in this paper, each SU is able to perfectly estimate the adapted modulation order (number of bits per modulation symbols) of its nearest PU link. Due to the use of AMC at every transmission link in the system, the relation between modulation order and the experienced SINR at $i$th. primary link receiver can be expressed as \cite{cioffinotes,b5}, 

\begin{eqnarray}\label{Modorder}
M_i^{(p)}=\log_{2} (1+k\gamma_i^{(p)}),\\ [-.2in] \nonumber
\end{eqnarray}
where $M_i^{(p)}$ is the $i$th. PU's modulation order, $k$ is the inverse SNR gap constant $k=\frac{1.5}{-\ln(5BER)}$ which depends on a target maximum transmit bit error rate (BER) requirement, and $\gamma_i^{(p)}$ is the SINR at the receiver of this link. During the listening stage, the $\gamma_i^{(p)}$ can be expressed as,
\begin{eqnarray}\label{sinrprimwsus0}
	\gamma_i^{(p)}[0] = \frac{G_{ii}^{(pp)} P_{i}^{(p)}[0]}{I_{pp}[0] + \sigma_p^2 },\\ [-.1in] \nonumber
\end{eqnarray} 
\noindent where $I_{pp}[0] = \displaystyle{\sum_{j=1,j{\ne}i}^{N_P}} G_{ij}^{(pp)} P_{j}^{(p)}[0]$ denotes the interference from the PN to the $i$th. primary link, and $\sigma_p^2$ is the background noise power. Using \eqref{Modorder} and \eqref{sinrprimwsus0}, $P_{i}^{(p)}[0]$ can be expressed as follows,
\begin{eqnarray}\label{primarypower0}
	P_{i}^{(p)}[0] = \frac{(2^{M_i^{(p)}[0]}-1)(I_{pp}[0] + \sigma_p^2 )}{k {G_{ii}^{(pp)}}},\\ \nonumber
\end{eqnarray} 
\noindent where $M_i^{(p)}[0]$ denotes the PU's modulation order. During the probing stage, due to the additional interference created by the SN, $\gamma_i^{(p)}[0]$ is changed to be:
\begin{eqnarray}\label{sinrprimwsus1}
	\gamma_i^{(p)}[1] = \frac{G_{ii}^{(pp)} P_{i}^{(p)}[1]}{I_{pp}[1] + \displaystyle{\sum_{j=1}^{N_S}} G_{ij}^{(ps)} P^{(s)} + \sigma_p^2 },\\ [-.1in] \nonumber
\end{eqnarray}

\noindent where $I_{pp}[1] = \displaystyle{\sum_{j=1,j{\ne}i}^{N_P}} G_{ij}^{(pp)}P_{j}^{(p)}[1]$ denotes the interference from the PN to the $i$th. primary link. Using \eqref{Modorder} and \eqref{sinrprimwsus1}, $P_{i}^{(p)}[1]$ can be expressed as:
\begin{eqnarray}
P_{i}^{(p)}[1] =\frac{(2^{M_i^{(p)}[1]}-1){\left (I_{pp}[1] + \displaystyle{\sum_{j=1}^{N_S}} G_{ij}^{(ps)}P^{(s)}+ \sigma_p^2 \right )}}{kG_{ii}^{(pp)}},\label{twocoleq1} \\ [-0.1in] \nonumber
\end{eqnarray}
\noindent where $M_i^{(p)}[1]$ denotes the PU's new adapted modulation order at the probing stage.
\begin{figure*}[!t]
\begin{eqnarray}
\frac{S_i[1]}{S_i[0]} =
\frac{{\Tilde{G}_{in}^{(ps)} \frac{(2^{M_i^{(p)}[1]}-1){(I_{pp}[1] + I_{ps} + \sigma_p^2) }}{k {G_{ii}^{(pp)}}}
+ \displaystyle{\sum_{\substack{j=1 \\ j{\ne}n}}^{N_P}}\Tilde{G}_{ij}^{(ps)}P_{j}^{(p)}[1]+I_{ss}}}{\Tilde{G}_{in}^{(ps)}\frac{(2^{M_i^{(p)}[0]}-1)){(I_p[0] + \sigma_p^2)}}{k {G_{ii}^{(pp)}}} +\displaystyle{\sum_{\substack{j=1 \\ j{\ne}n}}^{N_P}}\Tilde{G}_{ij}^{(ps)}P_{j}^{(p)}[0]},\label{twocoleq2} \\ [-.1in]\nonumber
\end{eqnarray}
\end{figure*}
By substituting $P_{n}^{(p)}[0]$ with \eqref{primarypower0}, and $P_{n}^{(p)}[1]$ with \eqref{twocoleq1}, equation \eqref{probing} can be rewritten as \eqref{twocoleq2}, \noindent where $I_{ps}= G_{in}^{(ps)}P^{(s)} + \displaystyle{\sum_{j=1,j{\ne}n}^{N_S}} G_{ij}^{(ps)}P^{(s)}$ displays the interference from the SN to the $i$th. SU's nearest primary link after probing (recall that cross-channel gain is denoted by $G_{in}^{(ps)}$).

In the particular case when the primary and secondary network have each only one transmitting link, \eqref{twocoleq2} can be simplified as:
\begin{eqnarray}
\frac{S_i[1]}{S_i[0]}= \frac{(2^{M_i^{(p)}[1]}-1)(G_{in}^{(ps)}P^{(s)}+\sigma_p^2)} {(2^{M_i^{(p)}[0]}-1)\sigma_p^2},\\ \nonumber
\end{eqnarray}\label{probing3}
Provided that the $\sigma_p^2$ is known by the SU, the ratio $\frac{S_i[1]}{S_i[0]}$ can be used to estimate the cross-channel gain and subsequently the interference from only SU link in the SN to the PU, as studied in \cite{b113} and \cite{b114} (note that $M_i^{(p)}[1]$ and $M_i^{(p)}[0]$ are known at the SU transmitter). Yet, in the general case we are considering here, where there are multiple transmission links in both networks, \eqref{twocoleq2} is the appropriate expression. However, \eqref{twocoleq2} is not of practical use because it requires from the SUs knowledge of too many system variables (the gains, etc.). Under the assumption of a fully autonomous network, it is not realistic to assume knowledge of all these involved parameters and to be able to apply a closed-form solution to this problem (and, specifically, being able to estimate the cross-channel gains that corresponds to each SU). Therefore, we posit the use of artificial neural network to solve the problem of SUs assessing their effect on the PN. The rational for this decision rests on the artificial neural network's property of being known as universal function approximator and their ability to implicitly extract, during the learning (training) process, the interrelation between the system variables. Leveraging these characteristics of artificial neural networks in our proposed technique allows for the SUs to assess their transmission effect on the PN without the need to calculate intermediate magnitudes (e.g. the cross-channel gain(s)).

\section{Leveraging Adaptive Modulation and Coding in Underlay DSA}\label{AMC}

Before presenting our proposed underlay DSA technique, in this Section we outline the main ideas on how AMC can be leveraged to address the two main challenges associated with underlay DSA: 1) How to establish the interference threshold in the PN and 2) how SUs can autonomously become aware of the interference they create on the PN. As previously noted, AMC (or, as also called, link adaptation) has been used during the past two decades in practically all high-performance wireless communications standards and, as such, is assumed to be used in both the primary and secondary networks in this work. In this section we summarize some of the main features of AMC that are relevant to our fully autonomous and distributed underlay DSA scheme.
	
\begin{figure}[hbp!]
		\centering
		\includegraphics[width=0.53\textwidth]{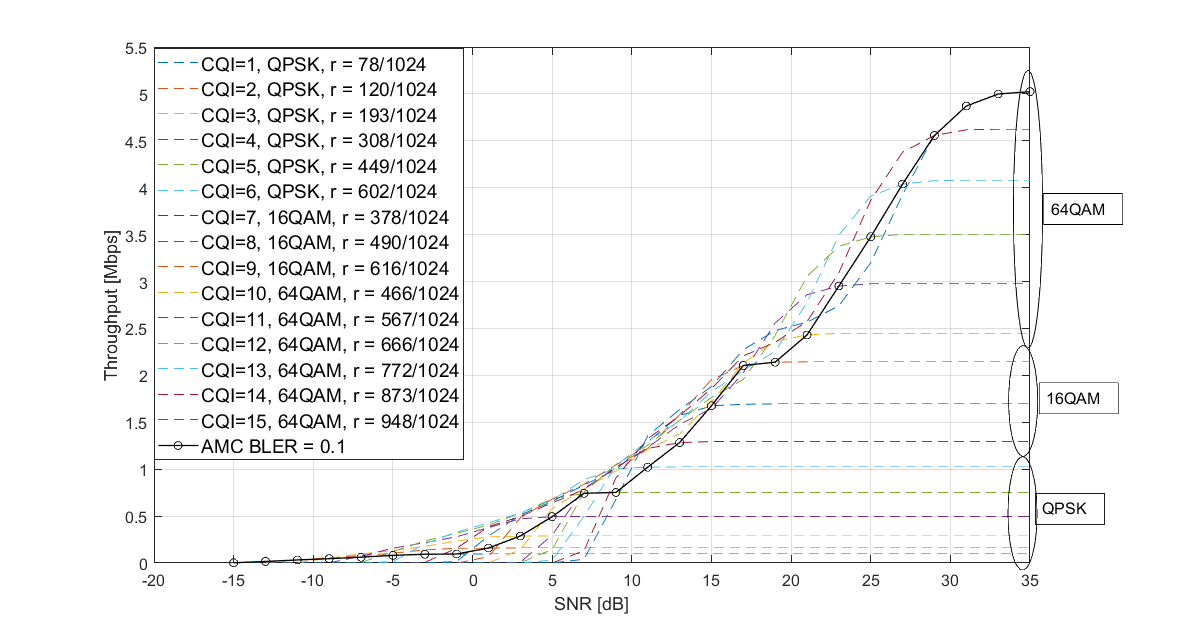}
		\caption{Throughput for each CQI and maximum throughput adaptive selection
        per RB for the AMC scheme of an LTE system under a Pedestrian B channel.}
		\label{LTEAMC}
	\end{figure}

Fig. \ref{LTEAMC}, obtained using the Matlab LTE Link Level Simulator from TU-Wien \cite{b21,IMT}, shows the throughput versus signal-to-noise ratio (SNR) performance for the LTE system under a Pedestrian B channel that will serve without loss of generality as the assumed AMC setup for the rest of this work (setup details for the simulation results shown in Fig. \ref{LTEAMC} are discussed in Section \ref{results}). In LTE, AMC consists of 15 different modes (each for a different ``Channel Quality Indicator" - CQI) based on three possible modulation schemes which will be called ``\emph{type 0}" for QPSK (used for the smaller SNR regime), ``\emph{type 1}" for 16QAM (used at intermediate SNRs), and ``\emph{type 2}" for 64QAM (used for the larger SNRs). In AMC, alongside the modulation order, channel coding rate is also adapted  \cite{b11,b12}. Fig. \ref{LTEAMC} shows the throughput of one LTE resource block achieved for each AMC mode (each curve is labeled with the corresponding CQI value and AMC mode settings, formed by the modulation type and channel coding rate) and the overall performance curve of the AMC scheme, where the modulation type and code rate are chosen to maximize throughput but with a constraint on the block error rate (BLER) not to exceed 10$\%$. During transmission, the transmitter chooses the AMC mode with maximum throughput at the estimated SNR of the link.
	
\begin{figure}[tbph]
		\centering
		\includegraphics[width=0.5\textwidth]{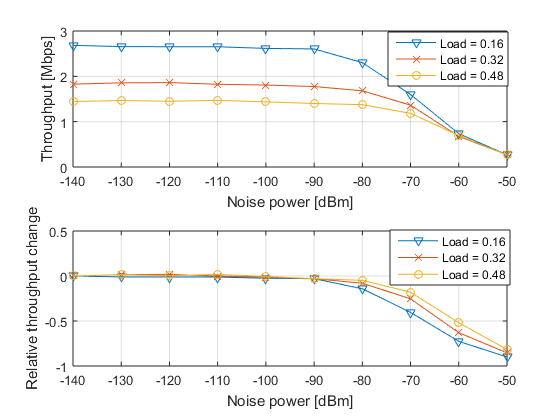}\vspace*{-3mm}
		\caption{Throughput vs. noise power in the primary network.}
		\label{trnoise}
	\end{figure}

As was discussed in \cite{b5}, the use of AMC in conjunction with transmit power control allows the background noise to increase up to a maximum value without significantly affecting the network average throughput. In the context of underlay DSA, this maximum noise value can be interpreted as the maximum value for the combined powers of background noise and interference from the SN. To see this important point in detail, consider Fig. \ref{trnoise}, which is an expanded version of Fig. 5 in \cite{b5}, now for different network loads ($N_P/N_{PBS}$) when using the LTE AMC setup just described and the power control algorithm from \cite{b5}. The figure shows the average throughput achieved in the PN by itself (the presence of an SN is not included in this result) as the background noise power increases. On the top, the figure shows the property associated with the use of adaptive modulation that for all network loads the average throughput remains approximately constant until noise power becomes sufficiently significant. This is not a trivial observation as networks with a higher load are operating in a regime more influenced by the interference rather than the noise, but it is clear from the results that adaptive modulation manages to maintain a balance between interference and noise-dominated operation. The bottom of Fig. \ref{trnoise} shows as a function of noise power, the change in throughput relative to the throughput at the lowest noise power. The result exposes the remarkable property that the interference that would be imposed by the SN, which can be considered by a PN that is unaware of the presence of another network as part of the background noise, will not significantly affect the average throughput in the primary network as long as the combined SN interference, the interference by other primary links and actual background noise remains below a threshold approximately equal to -85 dBm (although this number somewhat depends on the network load). Moreover, relative throughput change starts to decrease at approximately the same value of noise power for all network loads (around -90 dBm). This is a consequence of the link adaptation performed through AMC. Moreover, throughput relatively decreases faster with smaller network loads. We believe that this is because at smaller network loads, interference across the network is lower and a larger ratio of transmissions use the less resilient higher rate modulation types.
	
While AMC entails the adaptation of both modulation order and channel coding rate, it can be seen in Fig. \ref{LTEAMC} that the modulation order provides a coarse adaptation and that the channel coding rate enables a finer adaptation within each of the choices for modulation order. Moreover, an important difference between modulation order and channel coding rate adaptations is that while it is possible for a passive ``listener" of the AMC transmission to infer the modulation order through the use of modulation classification signal processing, it is not possible to infer the channel coding rate (or equivalently the full AMC mode). On the other hand, as can be seen in Fig. \ref{LTEAMC}, when the noise power increases (or equivalently the SN interference increases) the effect of AMC operation will be to change the AMC mode to one associated with a smaller CQI. At the same time, the modulation scheme with the smallest order is used for the smallest operating SINRs (because the transmission of less bits per symbol is more resilient to interference and noise). As the interference from secondary transmissions increases, primary links that are already using, in the absence of secondary transmissions, the lowest modulation order scheme will not switch to other modulation schemes because there is no other modulation scheme with fewer bits per symbols to switch to. This means that transmissions from an SU that otherwise would generate a change in modulation scheme would not result in any change when the nearest primary link is already transmitting with the modulation scheme with smallest bits per symbol. As a result, estimation of the modulation order provides for the SU transmitter with a control nob to infer the effect of SN transmission on the PN but with severe limitations due to the coarse information provided by the estimated modulation order. However, as seen in Fig. \ref{LTEAMC}, although the increase in interference to the PU (and particularly increase in SU transmit power) may not lead to change in modulation order, it indeed results in the change in channel coding rate. Thus, the estimation of the channel coding rate (and equivalently the full AMC mode) is necessary in order to be able to finely estimate the effect of SU transmissions. Indeed, there exists a large body of research in the area of modulation classification with some representative works briefly discussed in Sect. \ref{sec:intro} (e.g. \cite{b27,b16,b28}). Consequently, the techniques that existed before our work have been limited to use only the coarse information derived from the modulation order (e.g. \cite{b113,b114}) or to rely on the sharing of information between the PN and the SN through the SUs accessing the control feedback channel in the PN to learn the finer information associated with the channel coding rate used in a primary link (e.g. \cite{b118}). As will be seen, our proposed technique is able to overcome the limitation of inferring the channel coding rate without exchange of information between the PN and the SN and, as such, be able to use the fine-grained information provided by channel coding rate without the SN tapping into any control channel of the PN.
	
\begin{figure}[tbph]
		\centering
		\includegraphics[width=0.5\textwidth]{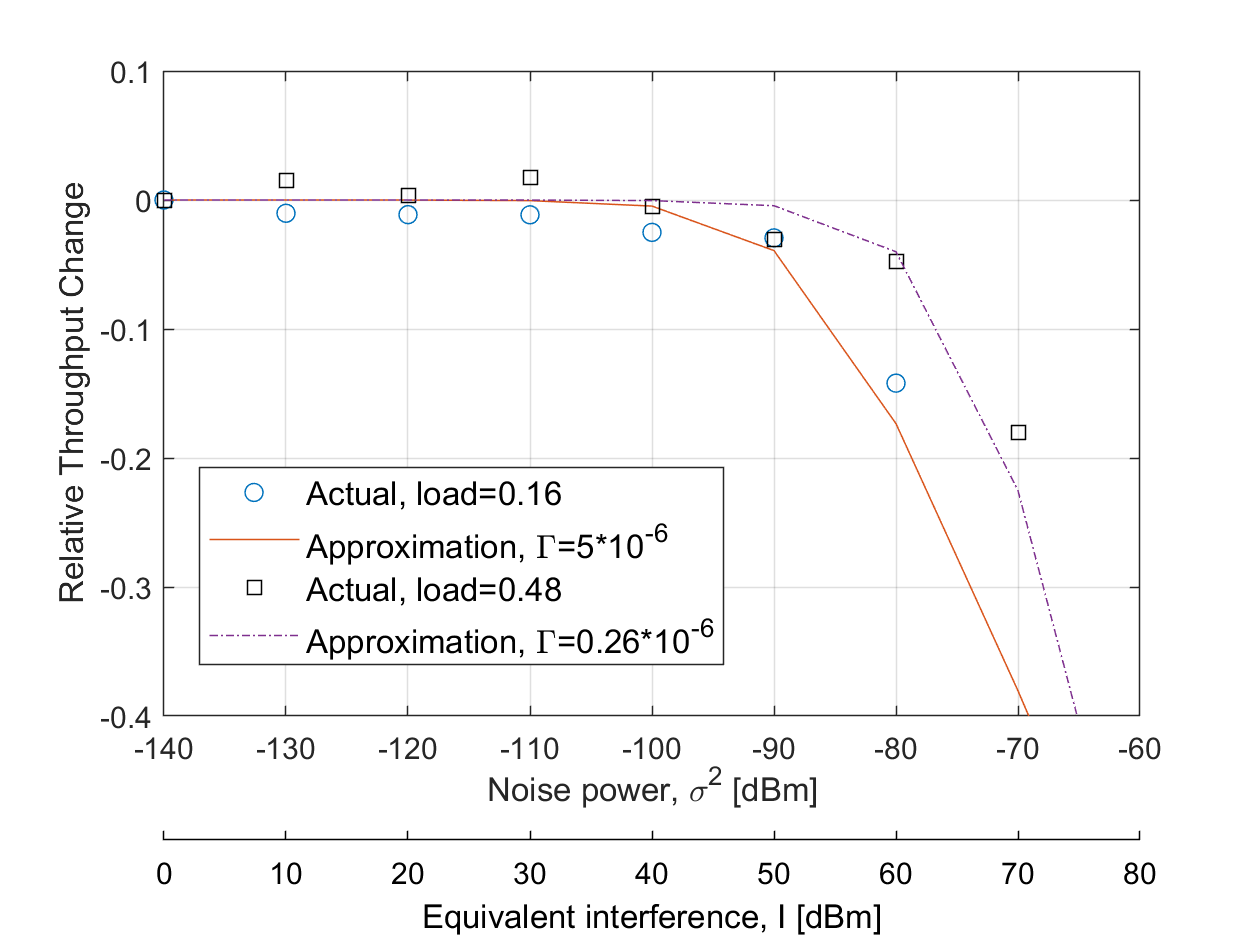}\vspace*{-3mm}
		\caption{Relative average throughput change and approximation \eqref{RelTPNApp}.}
		\label{fig_approximations}
\end{figure}

In the following we develop for a transmitting CR a model to estimate the interference limit from the PN. Regarding Fig. \ref{trnoise} it was stated that the property associated with the use of adaptive modulation allows the background noise
to increase up to a maximum value without significantly affecting the network average throughput; and that the maximum noise value can be interpreted as the maximum value for the combined powers of background noise and interference from the SN. For the purpose of underlay DSA, it is convenient to model the PN average throughput shown in the bottom of this figure as equal to the average throughput achieved without the presence of the SN minus an average throughput loss that is a function of an equivalent interference from the SN as experienced across the PN. This is, if $T_0(\sigma_p^2)$ is the PN average throughput without the SN (explicitly shown to be dependant on the background noise power $\sigma_p^2$), we model the PN average throughput $T_p$ as $T_p=T_0(\sigma_p^2)-T_l(I)$, where $T_l$ is the throughput loss and $I$ is the equivalent interference from the SN as experienced across the PN. Next, the throughput loss is modeled according to the SNR gap approximation for the Shannon's channel capacity formula, \cite{cioffinotes}, yielding
\begin{eqnarray}\label{TPNApp}
T_p=T_0 - B \log_2(1+\Gamma \, 10^{I/10}),
\end{eqnarray}
where $B$ is the PN system bandwidth, $I$ is the equivalent interference generated by the SN on the PN (measured in dBm), and $\Gamma$ is the SNR gap that accounts for the use of practical coding and transmission mechanisms (we have also embedded into $\Gamma$ a factor of 0.001 stemming from the conversion of units of $I$ from dBm). Being the relative average throughput change in the PN $T_{\%}=(T_p-T_0)/T_0$, from \eqref{TPNApp} we have,
\begin{eqnarray}\label{RelTPNApp}
T_{\%}=-\frac{1}{\zeta} \log_2(1+\Gamma \, 10^{I/10}),
\end{eqnarray}
where $\zeta=T_0/B$ is the PN spectral efficiency that is achieved without the SN's effect. Figure \ref{fig_approximations} compares the relative average throughput change from Fig. \ref{trnoise} for loads equal to 0.16 and 0.48 only for figure clarity and their approximations \eqref{RelTPNApp}. The approximation uses for the spectral efficiency calculation the same system bandwidth $B$ as in the results in Fig. \ref{trnoise} (180 kHz) and, of course, the values $T_0$ from Fig. \ref{trnoise} at noise power equal to -140 dBm. Fig.  \ref{fig_approximations} validates our model given by \eqref{TPNApp} and \eqref{RelTPNApp} by showing a good approximation for the relative average throughput change (the approximation shows less accuracy when the relative average throughput change grows beyond 40\% which is of no concern because these are values of relative average throughput change too large to be used in the practical operation of underlay DSA). More importantly, Fig. \ref{fig_approximations} includes a second abscissa to highlight the meaning of the equivalent interference in the context of underlay DSA where the PN uses AMC. As mentioned earlier, the curves in Fig. \ref{trnoise} show for a network that uses AMC the change in average throughput as the background noise power increases. In the context of underlay DSA, the absence of information exchange between the PN and the SN implies that, in principle, the PN is oblivious of the presence or not of the SN. Therefore, the PN would experience the interference from the SN as an increase in the background noise, which is what we identify as the equivalent interference $I$. This is illustrated in Fig. \ref{fig_approximations} with the second abscissa, showing the increasing equivalent interference with the same value as the background noise. In this way, for example, if the actual background noise is -140 dBm, an equivalent interference of $I=30$ dBm is performance-wise perceived by the PN as a total background noise of -110 dBm. 

The goal of the proposed underlay DSA mechanism at the SN is to find the transmit power at the SUs that results in an equivalent interference, denoted as interference limit $I_0$, that is as large as possible (to increase throughput at the SN) while the relative average throughput change in the PN remains below a limit that we will denote as $\epsilon$. From \eqref{RelTPNApp}, this maximum equivalent interference, measured in dBm and denoted as interference limit, can be expressed as a function of the limit $\epsilon$ as,
\begin{eqnarray}\label{limI}
I_0=10*\log \Big(\frac{2^{-\epsilon\zeta}-1}{\Gamma}\Big).
\end{eqnarray}
Using this expression, we have calculated, for example with $\epsilon=-0.05$, values of maximum equivalent interference equal to  51.3 dBm and 61.1 dBm for load equal to 0.16 and 0.48, respectively. The expression in \eqref{limI} provides an answer to one of the two main challenges in underlay DSA: the establishment of the interference threshold from the PN. However, the second challenge remains: how the SUs autonomously become aware of the interference they create on the PN, relative to the limit $I_0$. This challenge is compounded by our setup where the SUs can only use information they sense individually (the modulation scheme used at the nearest primary link) which, as explained earlier, leads to many of the parameters that would be needed for an analytical solution being not accessible to the SUs. To solve this challenge we propose the use of artificial neural networks because of their state-of-the-art status within techniques capable of learning the relation between the equivalent interference and the SUs' transmissions (that is implicitly represented within the training data).


\section{Autonomous and Distributed Underlay DSA for Cognitive Radio Networks}\label{thetech}

We now present the main contribution of this paper: a fully autonomous and distributed underlay DSA technique for a secondary CR network. 
The operation of the SN is fully autonomous and ad-hoc. This means that SUs do not rely on any exchange of information with the PN and with other SUs (other than between transmitter-receiver pairs) and that the transmission control algorithm in the SN needs is distributed. While there is no information exchange between primary and secondary networks, it is assumed that the SN has knowledge of the underlying timing operation in the PN because it knows the protocol used in the PN (through standard or publicly available information). Because of this, the SN has knowledge of the underlying timing operation in the primary network at a broad level (e.g. when frame transmission starts, etc.) so that CRs are able to sense and transmit at appropriate times. Detailed (symbol-level) knowledge of timing at the CRs is not required or assumed.
	
As stated, fully autonomous operation implies that the primary and secondary networks operate as being unaware of the other (except for the mild timing assumption), considering the other network transmissions as out-of-network interference akin to background noise. Then, when adding an underlay SN, the SINR at the receiver of the $i$th. PN link now becomes,
\begin{eqnarray}\label{sinrprimwsus}
	\gamma_i^{(p)} = \frac{G_{ii}^{(pp)} P_i^{(p)}}{\displaystyle{\sum_{j{\ne}i}^{N_P}} G_{ij}^{(pp)} P_j^{(p)} + \displaystyle{\sum_{j=1}^{N_S}} G_{ij}^{(ps)} P_j^{(s)} + \sigma_p^2 },\\ [-.1in] \nonumber
\end{eqnarray}
where $P_j^{(s)}$ is the transmit power from the $j$th. transmitting SU and $G_{ij}^{(ps)}$ is the path gain from a transmitting SU $j$ to a PU $i$. Likewise, it is assumed that transmissions on the SN also make use of AMC, which is configured based on the corresponding link SINR. For this, the SINR, $\gamma_i^{(s)}$, at the receiver of the $i$th. SN link is:
\begin{eqnarray}
	\gamma_i^{(s)} = \frac{G_{ii}^{(ss)} P_i^{(s)}}{\displaystyle{\sum_{j{\ne}i}^{N_S}} G_{ij}^{(ss)} P_j^{(s)} + \displaystyle{\sum_{j=1}^{N_P}} G_{ij}^{(sp)} P_j^{(p)} + \sigma_s^2 }, \label{sinrsec}\\ [-.1in] \nonumber
\end{eqnarray}

All transmissions are assumed to be over a quasi-static fading channel that is considered constant during a sensing and transmission period. This setup assumed for the SN is general, yet practical, as it is applicable to numerous wireless cognitive ad-hoc networks scenarios.
	
In the proposed underlay DSA technique, a cognitive engine at each SN transmitter leverages the use of artificial neural network to learn the functional model of the interaction between the secondary and primary networks.

Often, analytical models have been used to characterize the performance of the SN. For example, in \cite{b13} the BER performance of different modulation orders have been characterized using analytical models. However, as shown in section \ref{MathAnalsis}, our derived analytical model requires from SUs knowledge of many variables which necessitate the use of control channels between two networks (which violates our main assumption of having fully autonomous setup).
Black-box modeling is an alternative approach to analytical models which considers analyzing the relation between input and output of a system and aims at building a predictor to estimate output values for unforeseen inputs and variations of the system configurations. In this work we will follow a black-box modeling approach.
	
Neural Networks (NNs) have become increasingly popular as general purpose function approximators and, specifically, for dynamic system modeling \cite{b14}. Neural networks have been successfully applied to a number of black-box modeling and time series prediction tasks. Due to the inherent capability of neural networks in modeling nonlinear systems and their higher robustness to noise, they frequently outperform standard linear techniques when the time series are noisy and the dynamical system that generated the time series is nonlinear \cite{box2015time}. There is a growing number of works that have applied neural networks for various communication tasks such as channel decoding, estimating the features of the user channels and predicting the anomalies for wireless sensor networks \cite{lyu2017performance,navabi2018predicting,anomali2018predicting}. For CRs, the feed forward neural network has been used in predicting the spectrum occupancy status \cite{b9} and designing a medium access control (MAC) protocol \cite{yu2018deep}. 

Due to the adaptation of SU's transmission to the PN interference threshold, an underlay network can be seen as an example of a dynamic system, and also the throughput in the PU can be seen as a time series with a temporal dependency. As a result, we have considered a neural network-based cognitive engine to specifically predict the throughput in PU and characterize the behavior of such dynamic system. In the case of one-step-ahead time series prediction tasks, only the estimation of the next sample value of a time series is required. Therefore, the input contains only actual sample points of the time series, without feeding back the output as a new input to the model. While considering multi-step-ahead or long-term prediction, the neural network's output should be fed back to the model as a new input for a finite number of time steps \cite{sorjamaa2007methodology}. In this case, the components of this input to the neural network, previously composed of actual samples of the time series, are gradually replaced by previously predicted values. As a result, the multi-step-ahead prediction task is converted to a dynamic modeling task. In this case, the neural network model behaves as an autonomous system and tries to recursively emulate the dynamic behavior of the system that generated the nonlinear time series \cite{haykin1995detection}. Compared to the one-step-ahead prediction, multi-step-ahead prediction and dynamic modeling are much more complex to deal with. However, neural networks models and in particular recurrent neural architectures play an important role in dealing with these complex tasks \cite{principe2000neural}. Elman introduced in \cite{elman1990finding} a class of recurrent neural models called simple recurrent networks (SRNs) which are essentially feedforward in the signal-flow structure with a few local and/or global feedback loops. A time delay neural network (TDNN), which is an adapted version of a feedforward multilayer perceptron (MLP)-like networks with an input tapped-delay line, can be used to process time series \cite{principe2000neural}. In the case of long-term predictions, a feedforward TDNN model will eventually behave similarly to the SRN architecture, since a global loop is needed to feed back the current estimated value into the model's input. Temporal gradient-based variants of the backpropagation algorithm are usually used to train the aforementioned recurrent neural networks \cite{pearlmutter1995gradient}. However, training using gradient-based learning algorithms can be quite difficult in the case of systems with a long time temporal dependencies between their input-output signals \cite{bengio1994learning}. In \cite{lin1998embedded}, the authors claimed that such training is more effective in a class of simple recurrent network models called Nonlinear Autoregressive with eXogenous input (NARX) \cite{lin1996learning,b8,b18} than in simple MLP-based models. This class of neural networks were proven to be powerful in pattern recognition and classification applications as well \cite{b7,lin1997delay}. In this paper, the architectural approach proposed for modeling the underlay system was chosen based upon the NARX neural network structure. Specifically, we choose the NARX neural network to implement the cognition task at the cognitive engine in each SU.
	
\begin{figure}[tbp]
		\centering
		\includegraphics[width=0.45\textwidth]{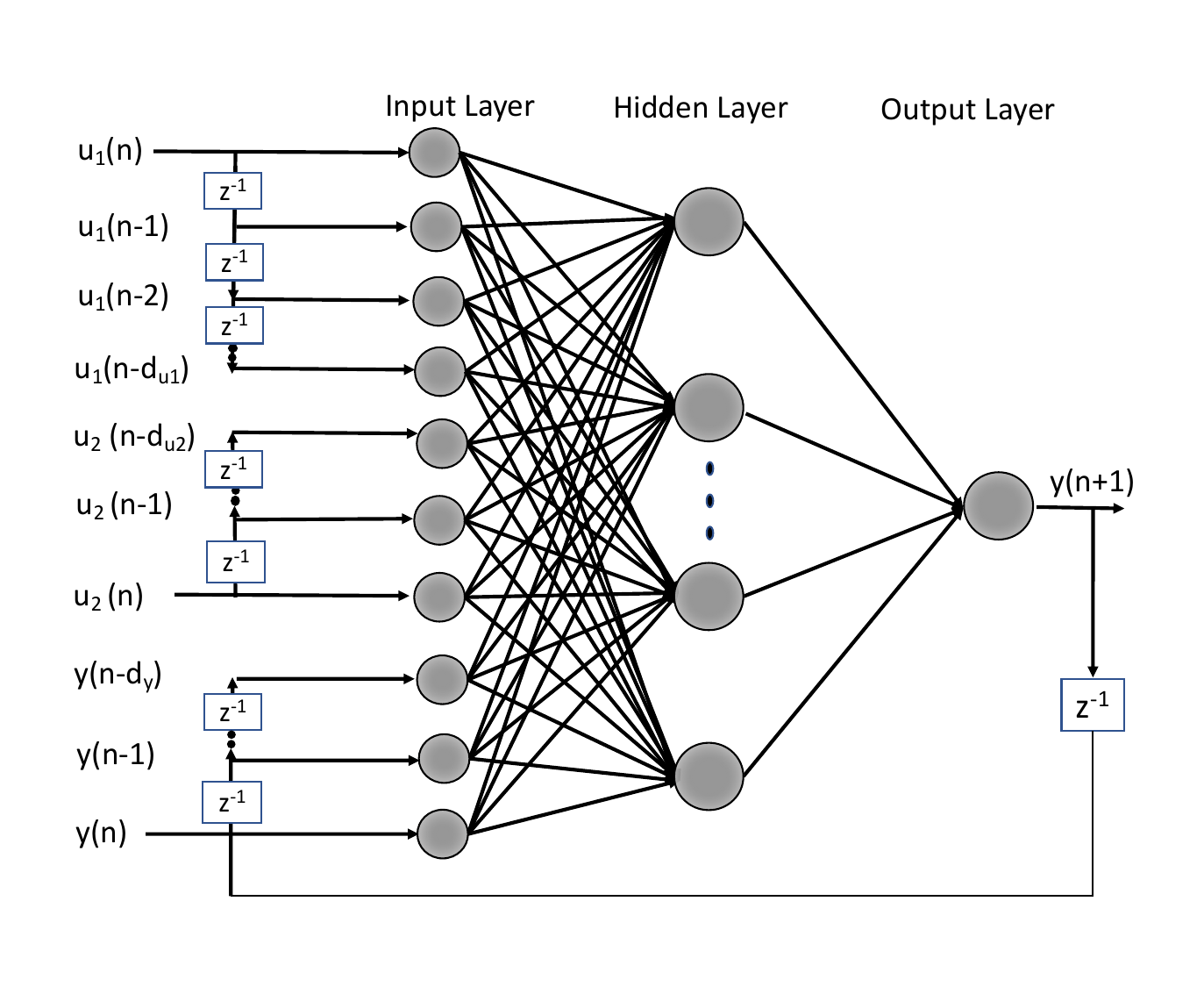}\vspace*{-1mm}
		\caption{NARX neural network with two delayed inputs and one delayed output.}
		\label{NARXNN}
\end{figure}

With a topology as shown in Fig. \ref{NARXNN} for the case of one hidden layer network, the NARX neural network output can be mathematically represented as \cite{b15},
\begin{eqnarray}
	\label{narx}
	y(n+1) =f\Big(y(n),y(n-1),\dots,y(n-d_y);\nonumber\\
	u_1(n),u_1(n-1),\dots,u_1(n-d_{u_1});\nonumber\\
	u_2(n),u_2(n-1),\dots,u_2(n-d_{u_2})\Big),\\ [-.2in] \nonumber
\end{eqnarray}
where $u_1(n)$, $u_2(n)$ and $y(n)$ denote, respectively, the two inputs and one output of the model at discrete time step $n$, and $d_{u_1}\geq1$, $d_{u_2}\geq1$ and $d_y\geq1$, $d_{u_1}\geq d_y$ , $d_{u_2}\geq d_y$ are the inputs and output discrete delays, respectively. The nonlinear mapping $f(\cdot)$ in \eqref{narx} can be approximated, for example, by a standard multilayer neural network. If the non-linear mapping can be learned accurately by a neural network of moderate size (measured in terms of number of layers and number of artificial neurons in each layer), the resource allocation based on the output of the NARX neural network can be done in real time, since passing the input through the neural network only requires a small number of simple operations. 
	
In Fig. \ref{NARXNN} each circle represents an ``artificial neuron'', an elementary operation unit in the NARX neural network model. According to the input variables $u_1(n)$ and $u_2(n)$, the output of $i$-th hidden layer neuron at time step $n$ is obtained as:	
\begin{eqnarray}
	\label{neuron1}
	H_i(n)=\phi_1\bigg(\sum_{r=1}^{d_{u_1}} w_{ir} u_1(n-r)+
	\sum_{k=1}^{d_{u_2}} w_{ik} u_2(n-k)+\nonumber\\
	\sum_{l=1}^{d_{y}} w_{il} y(n-l)+a\bigg),\\ [-.1in] \nonumber
\end{eqnarray}
\noindent where $w_{ir}$ is the connection weight between the input neuron $u_1(n-r)$ and $i$-th hidden neuron; $w_{ik}$ is the connection weight between the input neuron $u_2(n-k)$ and $i$-th hidden neuron; $w_{il}$ is the connection weight between the $i$-th hidden neuron and output feedback neuron $y(n-l)$; $a$ is the bias of the hidden layer (not explicitly shown in Fig. \ref{NARXNN}) and $\phi_1(\cdot)$ is called the ``activation function'' for the hidden layer. Combining the hidden layer output, the final prediction can be given as follows:
\begin{eqnarray}
	\label{neuron2}
	\hat{y}(n)=\phi_2\bigg(\sum_{i=1}^{n_{h}} w_{i} H_i(n)+b\bigg),\\ [-.1in]\nonumber
\end{eqnarray}

\noindent where $w_{i}$ is the connection weight between the $i$-th hidden neuron and predicted output; $n_h$ is the number of neurons in the hidden layer; b is the bias of the predicted output; and $\phi_2(\cdot)$ is the output layer activation function. In our implementation of the NARX neural netowrk, the activation function $\phi_1(\cdot)$ for the hidden layer is a sigmoid function while the activation function $\phi_2(\cdot)$ used for the output layer is linear (the input layer is not truly formed by artificial neurons but rather it is conventionally included as a representation of the connections of inputs into the neural network).
	
\begin{figure*}
		\includegraphics[width=0.95\textwidth,height=3cm]{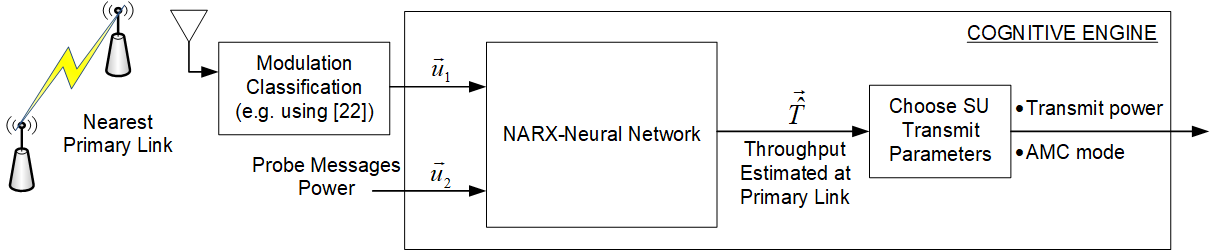}
		\caption{Block diagram of an SU with a cognitive engine based on the NARX neural network.}
		\label{CE}
\end{figure*}
		
 Fig. \ref{CE} illustrates the block diagram of an SU with its cognitive engine based on the NARX neural network, as well as other processing steps associated with the inputs and output of the neural network. The function of this cognitive engine will be the setting of power control and AMC parameters for a transmitting SU, by predicting for different transmit settings the throughput $\hat{T}$ (equivalently the CQI or AMC mode) of the nearest PN link. Under our imposed practical condition of no exchange of information between the primary and secondary networks, a CR can only estimate the modulation order (after performing modulation classification signal processing on the PN transmissions) and immediately has no direct way to know the coding rate in use. Moreover, practical limitations further dictate that the modulation classification can only be performed on the one primary transmission that it is being received with strongest power (usually this is the closest one), making the other transmissions be interference. It is assumed that the estimation of modulation type does not rely on the SU accessing any information from the PN feedback channel and is error free using any of the methods existing in the literature (e.g. \cite{b16} or \cite{b28}). As mentioned, the modulation order setting of a primary link provides a coarse indication of its experienced SINR, and accordingly of its experienced level of interference imposed by the transmissions in the SN. Since this created level of the interference has direct relation with the transmit power setting at each SU, we configure the NARX neural network cognitive engine to have as inputs the most recent values of assigned transmit power levels at the SU ($u_1(n)$) and, corresponding to these transmit power levels, the modulation order ($u_2(n)$) being used in the closest primary link that is inferred using a modulation classification technique. The output of the NARX neural network cognitive engine is the predicted throughput $\hat{T}(n)$ at the primary link closest to the transmitting CR. The predicted throughput on a primary link corresponds to a choice of AMC mode. 

\begin{figure}[tbph]
		\centering
		\includegraphics[width=0.45\textwidth]{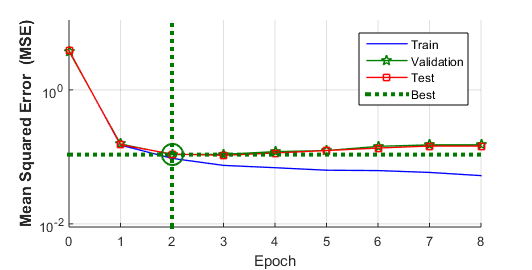}
		\caption{Throughput prediction performance of the NARX neural network. Best validation performance is 0.10825 at epoch 2.}
		\label{Train-performance}
\end{figure}

The proposed NARX neural network undergoes a training process to determine the values of its weights and biases. In this training process, the weight and bias values are updated according to the Levenberg-Marquardt optimization method. This method involves a back-propagation algorithm to compute the gradients of the prediction error corresponding to the artificial neurons \cite{mirzaee2009long}. It is known that in the case of function approximation problems, for the neural networks containing up to a few hundred weights, the Levenberg-Marquardt algorithm will have the fastest convergence \cite{diaconescu2008use}. Mean square error (MSE) was chosen as the prediction error metric:
\begin{eqnarray}
	\label{narxmse}
	MSE =\frac{1}{N} \sum_{i=1}^{N} (e_i)^2=\frac{1}{N} \sum_{i=1}^{N} (T_i-\hat{T}_i)^2,
\end{eqnarray}
where $N$ is the size of training data set, $T_i$ and $\hat{T}_i$ are target and predicted values, respectively. In order to find the number of hidden neurons and depth of the tap-delay lines in the NARX neural network, we perform a thorough set of experiments with different configurations. The best performing structure was found to be with 50 hidden neurons and 7 time delay steps. Fig. \ref{Train-performance} illustrates the performance of the NARX neural network for a primary network load equal to 0.64. Further details regarding the generation of the target prediction values $T_i$ used during training are provided in the next Section.
	
The underlay DSA operation of an SU based on the NARX neural network cognitive engine is as follows: First the SUs avoid transmission while the PN initially adjusts its power and AMC parameters using the iterative power control algorithm (\ref{PrPowIter}). During this stage which earlier was denoted as listening stage, the SUs listen to the PN transmissions and infer the modulation order used by their closest primary link. At this stage, each transmitting SU obtains an estimate of the modulation order used by their corresponding nearest primary link when the SN is not transmitting. Next, each SU proceeds to send a series of short probe messages configured with different transmit powers. To devise the new adapted modulation order, each SU performs modulation classification for each probe message on its nearest primary link. The sequence of probe message transmit powers and corresponding estimated modulation orders at the nearest primary link are fed to the NARX neural network, which provides the sequence of corresponding estimated throughput values at the nearest primary link, $\vec{\hat{T}}$. Note that this sequence of estimated throughput values includes the one with no transmissions from the SN. Since estimating the throughput is equivalent to estimating the CQI and the corresponding full AMC mode (channel coding rate and modulation order), the NARX neural network is able to provide an estimate of the nearest primary link SINR with a finer resolution than what could be derived from the modulation classification alone that is present at its input. The SU can use the sequence of estimated throughput values to infer what would be the effect of its transmission on the nearest primary link by comparing the change in throughput value against that without the SN transmission. As seen in Fig. \ref{CE}, the SU uses this information to find its own transmission parameters. We will consider two approaches for the SUs to choose transmit settings. In the first one, the SU chooses the maximum transmit power value that is estimated to keep the modulation order at its nearest primary link unchanged. In the second approach, the SU chooses the maximum transmit power value that keeps the relative change in throughput of its nearest primary link below a predetermined limit. Note that the second approach differs from the first one in that it fully uses the advantage provided by the NARX neural network in providing the finer resolution full AMC mode inference on the nearest primary link, instead of the coarser inference on modulation order that the first scheme is based on. Additionally, for both approaches, those SUs that estimate that their closest primary link is transmitting in the lowest rate AMC mode when the SN is not transmitting (because of already experiencing a very low SINR, likely leaving no room for added interference from the SN) are prevented from transmitting. This guarantees that the reduction in the average rate of the primary link experiencing the poorest channel quality (CQI equal to 1) is minimized.
	
Going back to Fig. \ref{LTEAMC} helps to gain an intuition into the NARX neural network cognitive engine operation. The NARX neural network receives as one input the possible SU's transmit power levels organized in sequence and, as another input, the corresponding modulation order sensed from the nearest primary link. During training, the NARX neural network learns to predict the throughput values at the nearest primary link that correspond to the two input sequences. Considering \eqref{sinrprimwsus}, we can think that the sequence of transmit power values that is input to the NARX neural network will yield a range of interference values, which will correspond to a ``segment" of SINR values in the abscissa of Fig. \ref{LTEAMC}. The position of this segment within the range of SINR values depends on the many factors reflected in \eqref{sinrprimwsus} (e.g. primary channel gain, interference from other SUs, etc.) but the NARX neural network has a sense of where the segment is thanks to the reference provided by the sequence of modulation orders at the input (e.g. if for the setup in Fig.  \ref{LTEAMC}, the sequence of modulation schemes are QPSK and 16QAM, the SINR segment is around 10 dB). During training, the NARX neural network is presented with multiple different such segments  from different wireless environment scenarios, eventually learning the throughput vs. SINR AMC performance curve. During operation of the NARX neural network (the testing phase in machine learning terms), sensing the sequence of modulation schemes that results from the sequence of probe messages with different power settings allows the NARX neural network to localize the segment of SINR values for the nearest primary link (in effect, finding the network scenario presented during training that best matches the existing wireless environment) and, consequently, predicts the corresponding throughput from the AMC performance curve.

\section{Simulation Results}\label{results}
	
The performance of the presented technique was evaluated through Monte Carlo simulations (150 runs) based on a PN ``playground'' consisting of a five-by-five BSs grid ($N_{PBS} = 25$) with neighboring base stations separated by a distance of 200 m. To avoid edge effects in the playground, the grid wraps around all its edges. One channel was singled out in the experiment and any of the base stations can have this channel active. In order to reflect realistic AMC settings, we assumed that all transmissions are based on an LTE 2x2 MIMO configuration, and that the channel of interest is one resource block with a bandwidth of 180 kHz.
	
As noted earlier, we assumed that there are $N_P$ active base stations that are using the same channel to communicate with their respectively assigned PU receivers. The location of the $N_P$ PU receivers is determined at random using a uniform distribution with the limitation that no base station could have more than one receiver assigned to it. Also, the receivers were connected to the base station from which they received the strongest signal. Transmit powers in the primary network were limited to the range between -20 and 40 dBm. The transmit power assignment for the $i$th. active primary link ($i=1,2,\dots,N_P$) follows the same algorithm as in (\ref{PrPowIter}). Each primary transmission considers the other network transmissions as out of network interference akin to background noise.

In the simulation, the SN consisted of $N_S = 4$ transmit-receive pairs of CRs, with the transmitters placed at random (also with a uniform distribution) on the PN playground, around their respectively assigned transmitter within a distance not exceeding 50 m. In the simulation setup we intended to reflect a situation where the PN had somewhat more capabilities (achieving larger throughput and communication range) than the SN because of being the incumbent to the spectrum band under consideration. Therefore, we assumed that the SUs were smaller devices with transmit power in the range of -30 to 20 dBm. Twenty equally spaced power levels in this range are considered as the set of allowed settings for transmission. The operation of the SN, as mentioned before, is fully autonomous and ad-hoc and the transmission control algorithm is distributed. It is also assumed that the SUs know the primary transmission protocol and are able to synchronize their operation with the primary transmissions.

All links assumed a path loss model for urban area given by $L = 128.1 + 37.6 \log d + 10 + S$ (in dBs), where $d$ is the distance between transmitter and receiver in km,  $S$ is the shadowing loss (modeled as a zero-mean Gaussian random variable with 6 dB standard deviation) and the penetration loss is fixed at 10 dB, \cite{b20}. Our assumed small scale fading follows a Pedestrian B model from \cite{IMT}. Noise power level was set at -130 dBm. It should be noted that all transmissions adopt AMC to adapt their transmission to the quality of their respective link.

Each neural network was trained$/$validated with a data set of 10000 samples collected from a network simulation with the setup just described. A subset of the data set (7000 samples or 70\% of the data set) was used to train the neural network and to present the CRs with new environmental conditions. The rest of data was used to compare the prediction performed by the trained neural network with the actual expected performance in order to encounter new environment conditions. In order to generate training data for the neural network cognitive engine, a comparable SN was devised that maintained the ability to use the estimated modulation order of the nearest primary link but without the cognitive engine shown in Fig. \ref{CE}. Instead, this comparable system implemented a distributed power control algorithm modified to incorporate the modulation order of the nearest primary link. A number of algorithms had been proposed for distributed power control in ad-hoc wireless network. One of the first ones, and the precursor to many related variants, is the Foschini-Miljanic algorithm, \cite{b29}, which implements an iterative distributed power control process so as to meet a target SINR. We adopted this iterative power allocation algorithm for the alternative SN that generated the data set to train the NARX-NN cognitive engine at each SU. Specifically, power is calculated for a secondary link $i$ at each iteration $m$ using the update formula,\vspace*{-1mm}
\begin{eqnarray}
	P^s_i[m+1] = \left(\frac{\beta_i}{\gamma^s_i[m]}\right)\: P^s_i[m],\\ [-.2in] \nonumber
\end{eqnarray}
where $P^s_i$ is the transmit power, $\beta_i$ is the target SINR and $\gamma^s_i[m]$ is the actual SINR measured in the $m$th. iteration which can be calculated through (\ref{sinrsec}). The fact that this algorithm associates power control with a target SINR in our case is a useful feature because the target SINR, when met, also determines the modulation order to be used as follows\cite{b5},
\begin{eqnarray}
	T_i^{(s)} &=&   \log_2(1+k \: \gamma_i^{(s)}), {\quad}i=1,2,...,N_S. \label{Tsec} \\ [-.2in]\nonumber
\end{eqnarray}
Consequently, we can think that instead of having a set of possible [modulation, channel code] pairs, now we have a set of target SINRs to choose from. Let $\mathfrak{B}$ $=\{b_1, b_2,\dots,b_K\}$ be this set, where target SINRs $b_i$'s are assumed to be sorted in ascending order. Of course, reducing the target SINR will result in decreasing the transmit power. As a consequence, the algorithm provides the mechanisms to both adapt transmit power and AMC settings in a distributed way. Moreover, for fair comparison to the system with the proposed NARX neural network cognitive engine, the transmit power from SUs also needs to be constrained by the goal to not degrade the SINR of the closest primary network link to the extent of reducing the modulation order (not having the NARX neural network, this SN we are comparing against cannot operate based on the use of throughput inferred for the nearest primary link and can only make use of the modulation order estimated from the modulation classification process). Modifying the Foschini-Miljanic algorithm by reducing the target SINR allows to manage this constraint by resulting in a reduction in the SU transmit power. As such, we adopted for the control of CR transmissions in the alternative SN this modified version of the Foschini-Miljanic algorithm, where the SU target SINR is progressively reduced until there is no change in the modulation order of the nearest primary link. We note here that while it is certainly possible to use one of the many existing enhancements to the Foschini-Miljanic algorithm, we chose to use the original version without improvements because its provides a baseline performance measure and because this power control algorithm or a variation of it are not the contribution of our work. Of course, the PUs in this system maintained the power allocation algorithm proposed in \cite{b5}, as explained in Sect. \ref{PNsetup}. We also note that this benchmark SN constitutes an implementation of the central principles of \cite{b113} while also managing practical considerations not addressed therein (e.g. multiple links in the SN and PN, distributed, ad-hoc operation of the SN, etc.), and, as such, serves the purpose of providing an indication of the performance improvements of our proposed technique versus prior works.

Moreover, as briefly mentioned earlier, note that in AMC the modulation order transmitting the smallest number of bits per symbol is used for the smallest operating SINRs (because it is more resilient to interference and noise). When the interference from the SN increases, the primary links that were already using the smallest possible modulation orders when there were no secondary transmissions will not switch to other modulation orders because there is simply no other modulation order with fewer bits per symbols to switch to. This means that transmissions from an SU that otherwise would generate a change in modulation order would not result in any change when the nearest primary link is already transmitting with the modulation order with smallest bits per symbol. Moreover, primary links using this modulation order, do so because their SINR is at the lower range of the operating SINRs, which implies that they are at a link state that likely may not leave much room for added interference from SUs. Because of these reasons, and in the interest of prioritizing the protection of primary links against excessive SN interference, we configured the alternative SN so that a CR will not transmit if it senses that its nearest primary link is using the lowest modulation order when the SN is not transmitting (as explained in Sect. \ref{thetech}, our proposed technique implements a mechanisms with the same spirit but based on checking for the smallest CQI at the nearest primary link when the SN is not transmitting, instead of smallest modulation order).
	
\begin{figure}[tbph]
		\centering
		\includegraphics[width=0.50\textwidth]{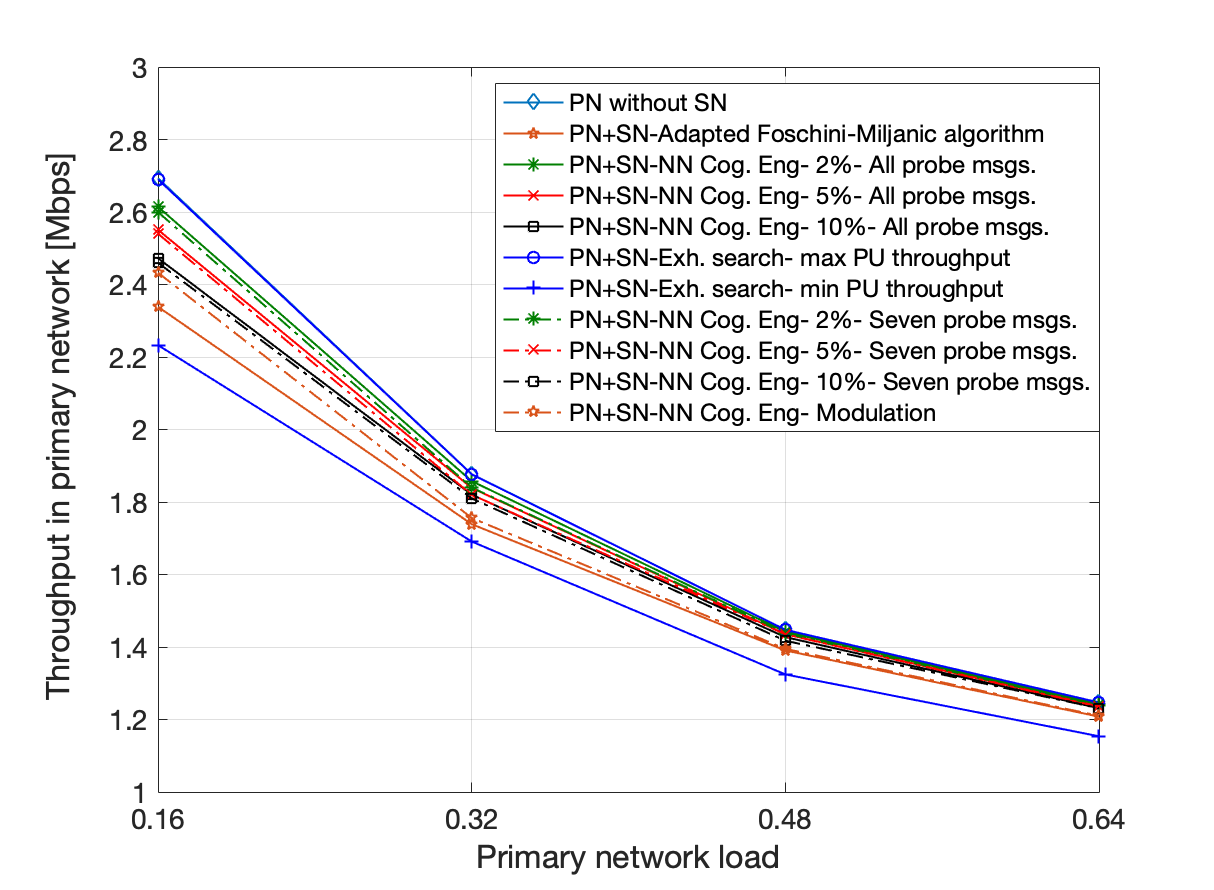}
		\caption{Average throughput in primary network.}
		\label{Tr-Load-PN}\vspace*{-1mm}
\end{figure}

\begin{figure}[tbp]
		\centering
		\includegraphics[width=0.5\textwidth]{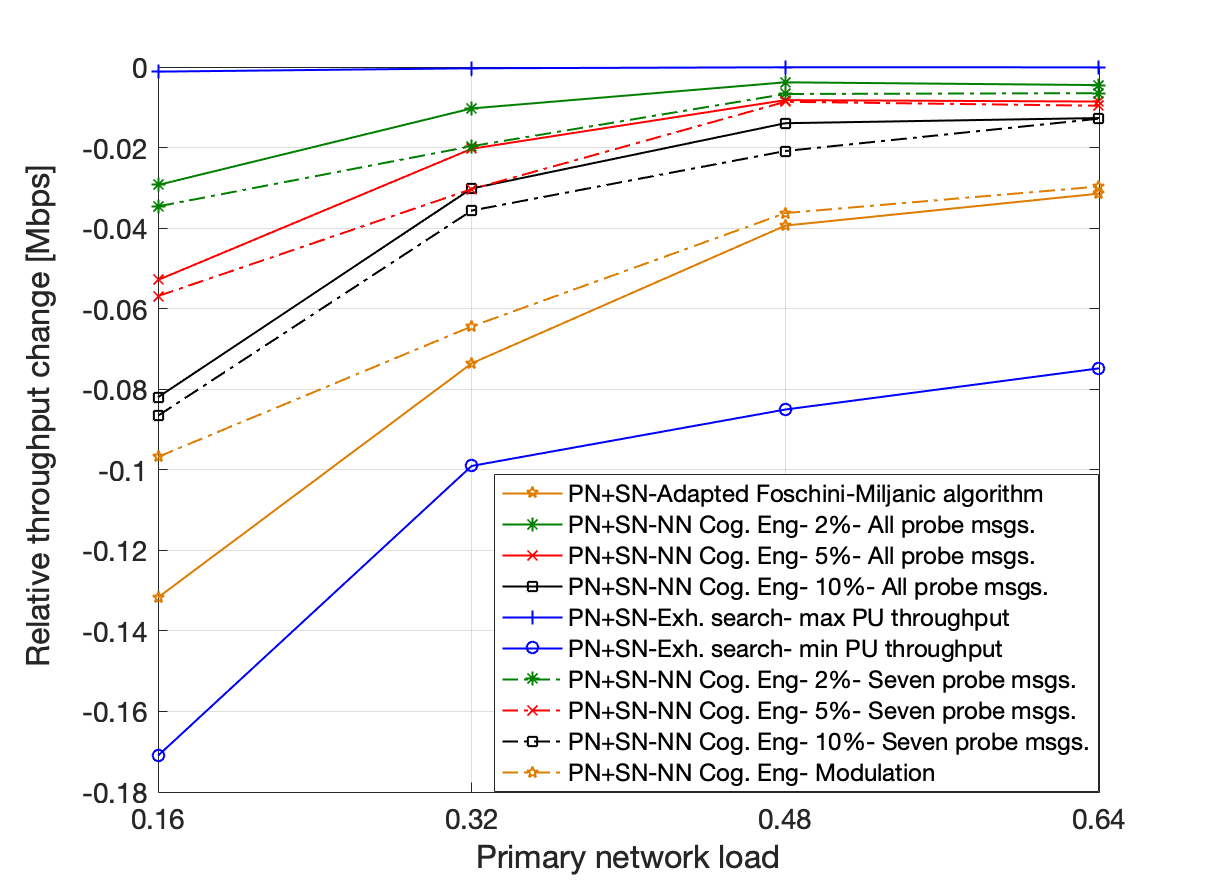}
		\caption{ Relative throughput change in primary network.}
		\label{Trchange-Load-PN}\vspace*{-1mm}
\end{figure}

\begin{figure}[tbph]
		\centering
		\includegraphics[width=0.5\textwidth]{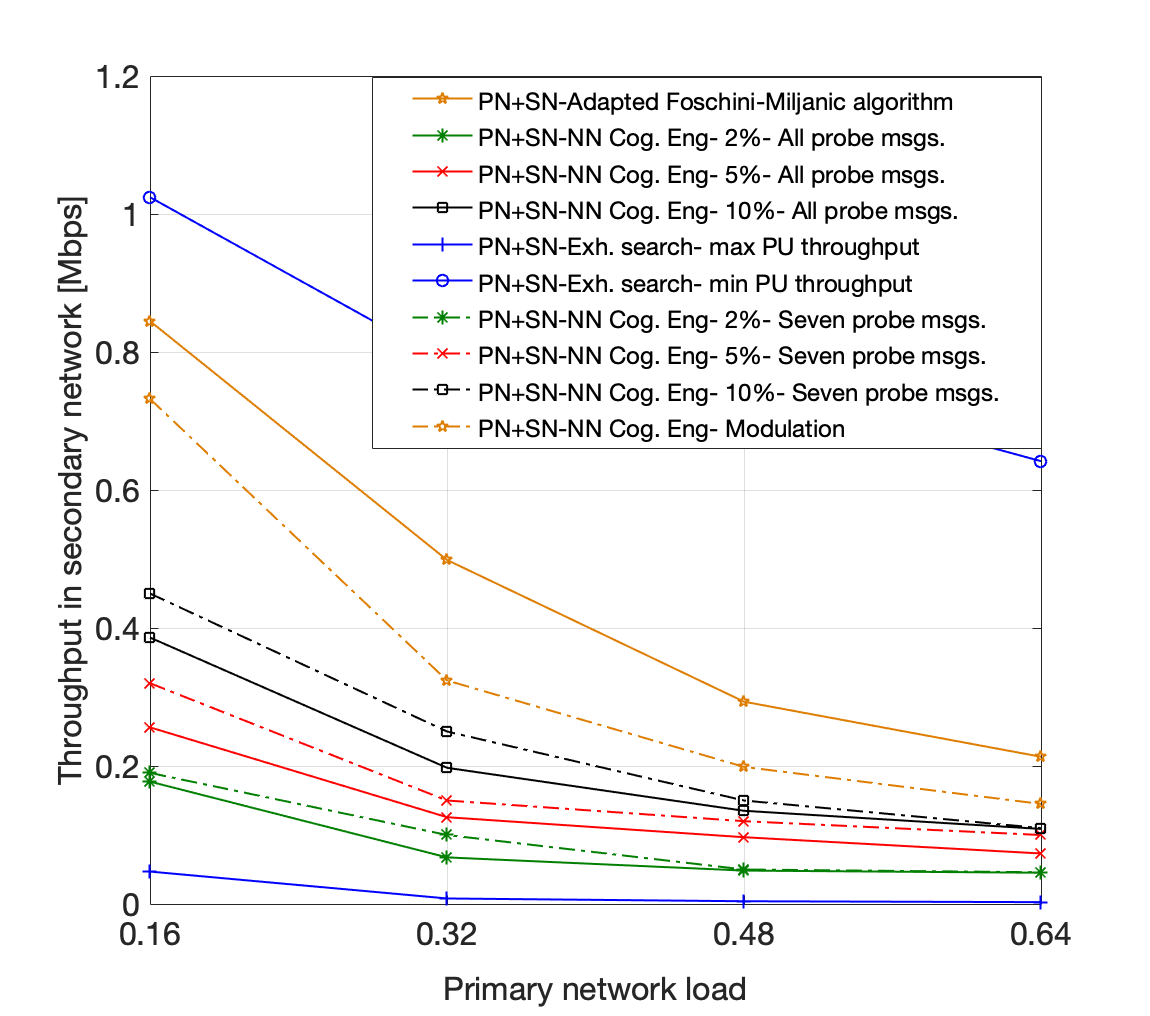}
		\caption{Average throughput in secondary network.}
		\label{Tr-Load-SN}
\end{figure}

Figs. \ref{Tr-Load-PN} through \ref{Tr-Load-SN} study the throughput performance in the primary and secondary networks for our presented technique and contrast them against other schemes. As indicated in Sect. \ref{thetech}, for the presented technique we considered two approaches for the SUs to choose transmit settings. The first approach, labeled in the figures as ``\emph{PN+SN- NN Cog. Eng.- Modulation}'', features our proposed cognitive engine with the capability for full AMC mode estimation at the nearest primary link, but it makes a limited use of this capability by making the SU choose the maximum transmit power value that is estimated to keep unchanged the modulation order (but not necessarily the channel coding rate) at its nearest primary link. The second approach makes full use of the cognitive engine's throughput (full AMC mode) estimation capabilities at the nearest primary link, by making the SU choose the maximum transmit power value that is estimated to not change the nearest primary link throughput beyond a maximum relative change value. This second approach is itself divided into a case where one probe message is sent for each possible power setting (for a total of twenty probe messages) and a second case that explores a reduction in the overhead from transmitting probe messages which transmits just seven messages and uses interpolation to complete the information for the rest of available transmit power settings. 
To show how our technique is able to leverage the estimation of the full AMC mode and provide each SU with a fine control over the interference it imposes to the nearest primary transmission link, we obtained results for three different limits on relative average throughput change in the PN: $2\%$, $5\%$ and $10\%$. In the figures, we labeled the results when sending all probe messages as ``\emph{PN+SN- NN Cog. Eng- 2\%- All probe msgs.}'' for a limit on relative average throughput change in the PN of $2\%$, ``\emph{PN+SN- NN Cog. Eng- 5\%- All probe msgs.}'' for a limit on relative average throughput change in the PN of $5\%$, and ``\emph{PN+SN- NN Cog. Eng- 10\%- All probe msgs.}'' for a limit on relative average throughput change in the PN of $10\%$. Similarly, for the case when sending seven probe messages, the labels for $2\%$, $5\%$ and $10\%$ limit on relative average throughput change in the PN are respectively ``\emph{PN+SN- NN Cog. Eng- 2\%- Seven probe msgs.}'', ``\emph{PN+SN-NN Cog. Eng- 5\%- Seven probe msgs.}'' and ``\emph{PN+SN- NN Cog. Eng- 10\%- Seven probe msgs.}''.  
	
The performance of these realizations of the proposed underlay DSA technique is compared in the figures against other schemes. The first such contrasting scheme, labeled in the figures as ``\emph{PN+SN- Adapted Foschini-Miljanic}'', is the same system used to collect training data and described earlier in this Section as the alternative SN. Recall that this scheme 
lacks the NARX neural network cognitive engine's capability to predict the full AMC mode at the nearest primary link. In addition, we considered two schemes that select transmit power for the SUs based on an exhaustive search across all possible setting permutations. In contradiction with our goal to avoid any exchange of information between the primary and secondary networks, these two schemes also incorporate the ability to perfectly know the CQI on the primary links as if the SN had access to the control feedback channels in the PN. 
The first exhaustive search scheme, labeled ``\emph{PN+SN- Exh. search-Max PU throughput}'', finds across all possible SUs transmit power permutations, the setting that results in no change in modulation order at any primary link and maximum average throughput in the PN. The second exhaustive search scheme, labeled ``\emph{PN+SN- Exh. search-Min PU throughput}'', favors the average throughput at the SN by finding across all possible SUs transmit power permutations, the setting that results in no change in modulation order at any primary link and minimum average throughput in the PN. Clearly, the two exhaustive search curves present extreme performance results based on ideal setups. Finally, Fig. \ref{Tr-Load-PN} includes a curve, ``\emph{PN without SN}'', which shows the average throughput achieved by the PN when the SN is not present.
	
Fig. \ref{Tr-Load-PN} shows the average throughput achieved for the PN as a function of the PN load, $N_P/N_{PBS}$, while Fig. \ref{Trchange-Load-PN} shows the change in this throughput relative to the ``PN without SN" case. It can be seen in both figures that the use of the proposed NARX neural network cognitive engine results in SUs transmit settings that reduces the average throughput in the PN much less than the case when the modified Foschini-Miljanic algorithm is used in the SU. Moreover, in the figures, the ``PN+SN- Exh. search-Min PU throughput'' curve illustrates the extent to which the average throughput in the PN can be affected without changing the modulation order at the nearest primary links (as much as 17\%). This indicates that considering only the modulation order provides an initial means for implementing a fully autonomous underlay DSA but with the limitations associated with the coarse indication of the primary links SINR given only by the modulation order. The ``PN+SN- Adapted Foschini-Miljanic'' and the ``PN+SN- NN Cog. Eng.- Modulation'' schemes, which both suffer from the limitation of relying on considering modulation order only, show better performance because SU transmit power is chosen in a more conservative way in terms of reducing effects to the primary network, instead of conducting an exhaustive search for the setting that results in minimum average throughput in the PU. Nevertheless, because of relying on modulation order inference only, the ``PN+SN- Adapted Foschini-Miljanic'' and the ``PN+SN- NN Cog. Eng.- Modulation'' schemes result in relative reduction in the PN average throughput by as much as 13.5\% and 9.5\%, respectively. The best performance in terms of controlling the effect of SN transmissions on the PN is achieved with our proposed scheme using the inference of the primary links' full AMC mode (in actuality, the throughput) provided by the NARX neural network cognitive engine. This is seen through the results obtained for the two cases (transmitting either all or seven probe messages) designed on the premise of limiting the maximum relative change in average throughput at the nearest primary link, for which we show results for 2\%, 5\% and 10\% relative change limit. Moreover, these schemes include the means to control as desired the level of SN effect on the PN (by setting the limit maximum relative change). In fact, the ``PN+SN- NN Cog. Eng- 2\%- All probe messages'' along with ``PN+SN- NN Cog. Eng- 2\%- Seven probe messages'' curves exemplify the very fine level of control that is possible to achieve with the proposed approach. 

Fig. \ref{Trchange-Load-PN} shows that the very fine level of control seen with our proposed scheme is achieved at all primary network loads, except at the lowest value of 0.16, when the relative change in average PN throughput exceeds the 2\% limit in only 1\% when sending all probe messages and in 1.5\% when sending seven probe messages. For the rest of cases, only in the case of 5\% limit, sending seven probe messages, and at the lowest primary network load, the relative change in PN average throughput is exceeded by just 0.5\%. 
These differences are attributed to errors in estimating the throughput at the PN, which are discussed more in detail later in this Section. Also, the differences are only seen at the smallest PN network load of 0.16 because of the larger sensitivity of the relative change in PN average throughput with lower PN load as was previously highlighted for Fig. \ref{trnoise}. Fig. \ref{Trchange-Load-PN} also shows that the scheme where a fraction of probe messages is used yields significant reduction in the transmission overhead of probe messages (roughly a threefold reduction) without much sacrifice in performance (maximum 3.5\% instead of 2\% actual achieved relative change in PN throughput compared to 3\% maximum change with all the probe messages, and maximum 5.5\% instead of 5\% actual achieved relative change in PN throughput compared to 5\% maximum change with all the probe messages). Finally, the curve ``PN+SN- Exh. search-Max PU throughput'' coincides with the ``PN without SN'' curve as the exhaustive search solution that maximizes average PN throughput is essentially the one with no transmissions in the SN (with one caveat to be discussed in Fig. \ref{Tr-Load-SN}).

Considering Fig. \ref{Trchange-Load-PN} in the context of the bottom plot of Fig. \ref{trnoise}, we can see that the relative throughput change in the PN when using our proposed technique corresponds to points at equivalent background noise power between -96 and -85 dBm (depending on the target maximum PN relative throughput change setting) just to the left of the ``elbow" of the curves in Fig. \ref{trnoise}. Equivalently, in the context of Fig. \ref{fig_approximations}, the related equation \eqref{limI}, and also depending on the preset maximum allowed change in PN relative throughput our proposed technique leads to equivalent interference to the PN between 44 and 55 dBm. This result confirms that our proposed technique succeeds in its main goal of autonomously and distributively determining the transmit power of the SUs such that the interference they create remains below the maximum threshold value associated with the background noise power levels that a PN with adaptive modulation can sustain. At the same time, the proximity to the ``elbow" of the curves in Figs. \ref{trnoise} and \ref{fig_approximations} of the equivalent interference levels generated by the SN, indicates that our proposed algorithm is able to find transmit settings for the SUs that will result in as large SN throughput as could be allowed by the PN interference limit.
	
Fig. \ref{Tr-Load-SN} shows the average throughput achieved in the SN as a function of the PN load. Naturally, the more a scheme affects the PN throughput, the larger the SN throughput it could achieve. As such, it can be seen that our approach based on a limit maximum PN relative throughout change not only provides the means to control how much the PN is affected by the SN, but also it allows to control how large the average throughput at the SN is desired to be (at the expense of the PN). Even so, the realization with the more restrictive setting for the SN (the one with a maximum PN relative throughput change of 2\%) still achieves useful average throughput values between 180 and 50 kbps for a channel with 180 kHz bandwidth (in case of transmitting seven probe messages the average throughput at the SN is slightly larger which is consistent with the results in Fig. \ref{Trchange-Load-PN}).  Also, note that at low PN loads, the ``PN+SN- Exh. search-Max PU throughput'' system shows throughput values that imply transmission in the SN. This does not contradict our earlier statement that this result essentially coincides with the ``PN without SN'' case. Instead, the transmissions in the SN that are seen in this case correspond to infrequent setups where the SUs are located so far away from the few active primary links (consider that this effect occurs only at very low PN loads) that they can transmit with very low power with no practical effect on the PN.

\begin{figure}[tbph]
		\centering
		\includegraphics[width=0.5\textwidth]{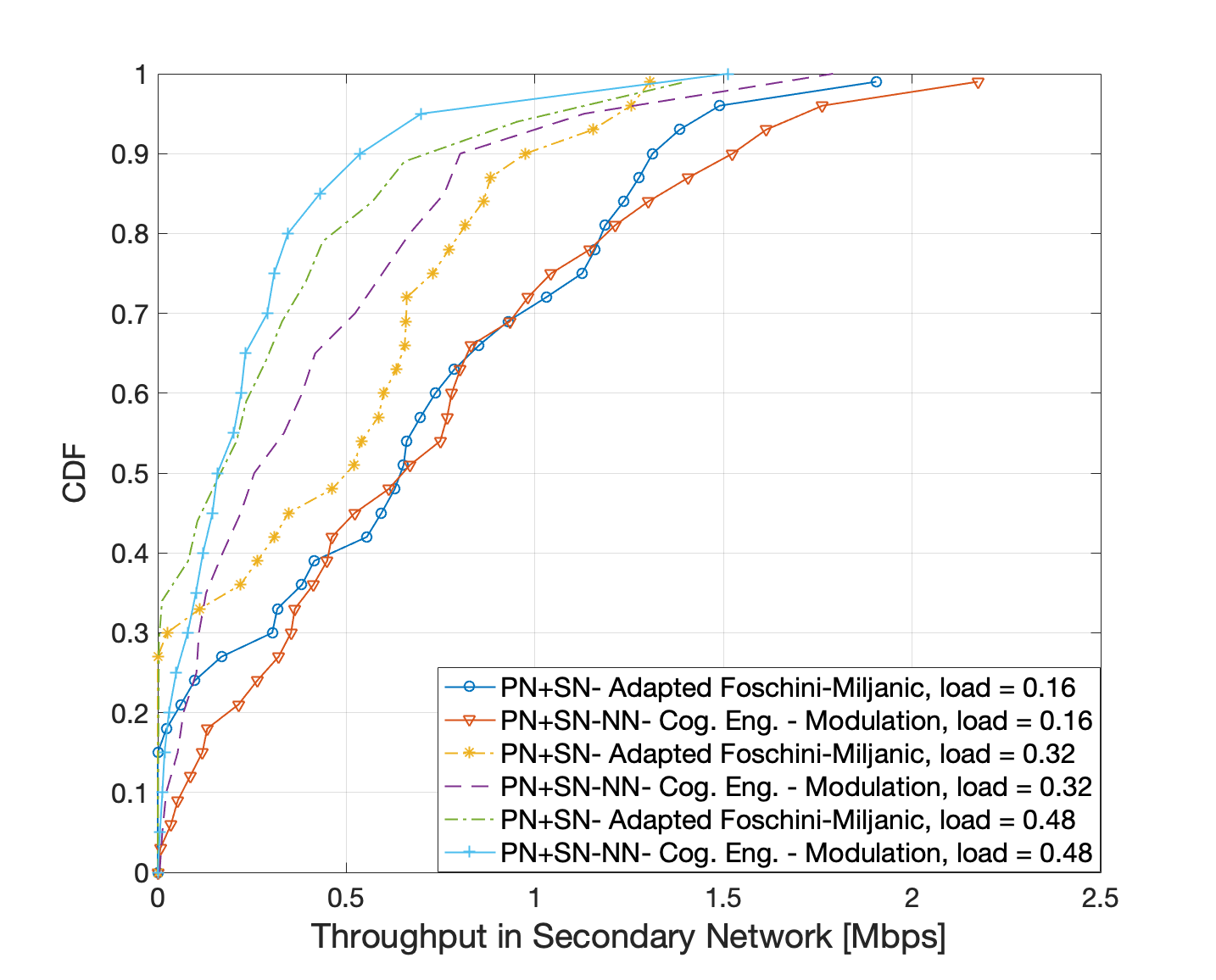}
		\caption{Cumulative distribution function (CDF) of the throughput in the secondary network.}
		\label{Tr-cdf-SU}
\end{figure}

Fig. \ref{Tr-cdf-SU} depicts the cumulative distribution function (CDF) of the throughput in the SN for different primary network loads. This figure presents a perspective that explains an added advantage of the NARX neural network solution compared to the modified Foschini-Miljanic algorithm-based solution. The figure shows that in the case of the SN that uses the modified Foschini-Miljanic algorithm, around 15\% of the time SUs will be unable to transmit (throughput is zero) when the PN load equals 0.16 and this number increases to around 30\% as the PN load increases. This is because the SN that uses the Foschini-Miljanic algorithm is only able to infer the modulation scheme used in the primary link and not the channel coding rate, which leads to SUs not being able to have a finer assessment of their effect on the PN when the nearest primary link is using a modulation ``\emph{type 0}". As a result, and as discussed earlier, in order to protect the PN, those SUs using the modified Foschini-Miljanic scheme for which the nearest primary link use modulation ``\emph{type 0}" are blocked from transmitting. In contrast, the proposed technique using NARX neural network, ``\emph{PN+SN-NN Cog.Eng.- Modulation}'', is able to estimate the finer AMC configuration of coding rate setting, making this protection and the blocking of SU unnecessary (except when the nearest link is using the AMC mode for a lowest rate, which corresponds to CQI=1). Consequently, as seen in Fig. \ref{Tr-cdf-SU}, the proposed technique increases the transmission opportunities in the SN by the same percentage of time that the SUs are blocked in the case of using the modified Foschini-Miljanic algorithm-based solution.
	
\begin{figure}[tbph]
		\centering
		\includegraphics[width=0.5\textwidth]{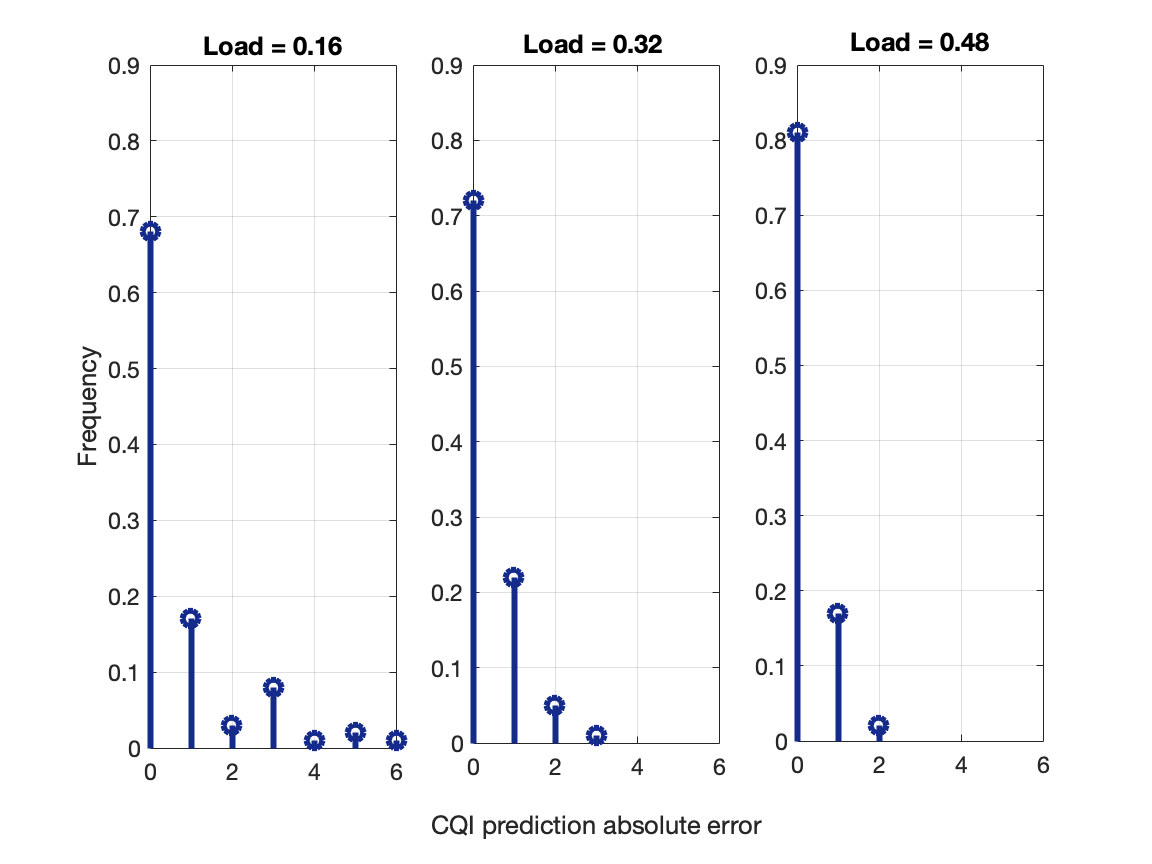}
		\caption{CQI estimation performance of the NARX neural network.}
		\label{CQI-performance}
	\end{figure}

As just seen, a key advantage of the proposed technique follows from the remarkable ability to estimate the channel coding rate used in a primary link. Therefore, we evaluated the performance of the NARX neural network in estimating the CQI in a primary link (which is equivalent to the full AMC mode consisting of modulation order and channel coding rate). Fig. \ref{CQI-performance} shows as a function of the primary network load the relative frequency of the absolute error when predicting the CQI for the case of transmitting all probe messages. This Figure shows that as the network load increases, the probability of an accurate estimation (prediction absolute error equal to zero) increases and reaches more than $80\%$ for a load equal to 0.48. The Figure shows that, overall, the probability of significant errors when predicting CQI is quite small but, nevertheless, we speculate this to be a factor in the (still small) reduction in PN average throughput and in the small difference at low PN loads between the target maximum relative change in average PN throughput and the actual achieved relative change in PN throughput for our schemes based on target maximum PN relative throughput change.
	
\section{Conclusion}\label{conclusion}

In this paper we have presented a fully autonomous and distributed underlay DSA technique that is based on a NARX neural network cognitive engine. The NARX neural network of a transmitting secondary network node learns to use the sensed modulation used in the nearest primary link to predict the effect of its transmission on the nearest primary network link. It does this by predicting the throughput or, equivalently, both the modulation scheme and the channel coding rate resulting from the configuration of adaptive modulation and coding at the nearest primary link. Based on this NARX neural network capability, we presented two variants for the proposed underlay DSA mechanism: one inspired in the current state of the art, where the SUs choose the maximum transmit power value that is estimated to not lead to a change in modulation order at their respective nearest primary link, and a second, more capable mechanism, where the SUs choose the maximum transmit power value that is estimated to not change their respective nearest primary link throughput beyond a chosen maximum relative change value. The performance of the latter proposed underlay DSA mechanism was examined for the cases of sending all or a third of all probe messages.  
	
Simulation results show that the proposed technique is able to accurately predict the modulation scheme and channel coding rate used in a primary link without the need to exchange information between the PN and the SN (e.g. access to feedback channels). Simulation results also demonstrated that the proposed technique succeeded in its main goal of determining the transmit power of the SUs such that their created interference remains below the maximum threshold that the primary network can sustain with minimal effect on the average throughput, along with reducing the transmission overhead when sending a fraction of probe messages. At the same time, it was seen that our proposed algorithm is able to find transmit settings for the SUs that will result in as large throughput in the SN as could be allowed by the primary network interference limit. Specifically, for a target PN maximum relative average throughput change of 2\% the proposed scheme is able to maintain the PN relative throughput change less than 3\% when sending all probe messages, and also less than 3.5\% when reducing three times the number of transmitted probe messages, while at the same time achieving useful average throughput values in the secondary network between 180 and 50 kbps for a channel with 180 kHz bandwidth. We also discussed how the ability of our proposed technique to predict the full AMC mode (not just the modulation scheme) results in a significant increase in the transmission opportunities in the SN compared to schemes that only use the modulation classification information.
	
\bibliographystyle{IEEEtran}
\bibliography{access}\vspace*{-20mm}
\end{document}